\pdfoutput=1 
\documentclass[aps,prb,showpacs,twocolumn,floatfix,longbibliography]{revtex4-2}

\usepackage{amsmath,amssymb,amsfonts}
\usepackage{graphicx,epsf}
\usepackage{xspace}
\usepackage{textcomp}
\usepackage{color}
\usepackage[normalem]{ulem}
\usepackage{fancyvrb}

\usepackage{listings}
\usepackage{xcolor}

\usepackage[T1]{fontenc}

\makeatletter
\g@addto@macro\@verbatim\small
\makeatother

\usepackage[htt]{hyphenat}

\usepackage[colorlinks,bookmarks=false,citecolor=blue,linkcolor=blue,urlcolor=blue,hyperfootnotes=true]{hyperref}

\usepackage{enumitem}
\usepackage{algpseudocode}
\usepackage{algorithmicx}
\usepackage{varwidth}

\newcommand{\be}{\begin{equation}}
\newcommand{\ee}{\end{equation}}

\newcommand{\sgn}{\text{sgn}}

\definecolor{codegreen}{rgb}{0,0.6,0}
\definecolor{codegray}{rgb}{0.5,0.5,0.5}
\definecolor{codepurple}{rgb}{0.58,0,0.82}
\definecolor{backcolour}{rgb}{0.95,0.95,0.92}
\definecolor{gray}{HTML}{f4f4f5}
\definecolor{warmgray}{HTML}{888888}
\definecolor{turquis}{HTML}{76b7b2}
\definecolor{orange}{rgb}{255,164,0}

\lstdefinestyle{mystyle}{
    backgroundcolor=\color{gray},
    keywordstyle=\color{black},
    numberstyle=\tiny\color{codegray},
    basicstyle=\ttfamily\scriptsize,
    columns  = fullflexible,
    breakatwhitespace=false,
    breaklines=false,
    captionpos=b,
    keepspaces=true,
    numbers=left,
    numbersep=5pt,
    showspaces=false,
    showstringspaces=false,
    showtabs=false,
    tabsize=2
}

\begin{document}

\title{Propagation of ultrashort voltage pulses through a small quantum dot}

\author{Thomas Kloss}
\email{thomas.kloss@neel.cnrs.fr}
\address{Universit\'e Grenoble Alpes, CNRS Grenoble, Grenoble INP, Institut N\'eel, 38000 Grenoble, France}
\author{Xavier Waintal}
\email{xavier.waintal@cea.fr}
\address{Universit\'e Grenoble Alpes, CEA, Grenoble INP, IRIG, PHELIQS, 38000 Grenoble, France}

\begin{abstract}
The coherent transport of time-resolved ultrafast excitations in nanoelectronic interferometers is expected to exhibit an interesting interplay between the interferences and the time-dependent drive.
However, the typical frequencies required to unlock this physics are in the THz range, making its observation challenging.
In this work, we consider the propagation of the excitation generated by ultrashort voltage pulses through a small quantum dot,
a system which we argue can display similar physics at significantly lower frequencies.
We model the system with a single resonant level connected to two infinite electrodes subjected to a time-dependent voltage bias.
For short pulses, we predict that the behaviour of the dot contrasts sharply with the long pulse (adiabatic) limit: the current actually oscillates with the amplitude of the voltage pulse. In the ultrafast limit, we predict
that the current can even be negative, i.e.\ flow against the voltage drop.
Our results are obtained by a combination of two approaches that are in quantitative agreement: explicit analytical expressions in
the ultrafast and ultraslow limits and exact numerical simulations. We discuss the applicability of our findings and conclude that this system should be within reach of existing experimental platforms.
\end{abstract}

\date{June 4, 2025}
\maketitle



\section{Introduction}

\label{sec:intro}
An interesting trend in quantum nanoelectronics  is to probe coherent devices with increasingly short voltage pulses. 
In particular Lorentzian shaped pulses, also known as Levitons \cite{levitov96, *levitov97, *keeling06}, 
allow the injection of single electron excitations without perturbing the Fermi sea \cite{Dubois13, Jullien14}. Levitons have also been proposed to serve as a novel form of quantum bits, also known as electronic flying qubits \cite{Bauerle18}.
In practice, the devices are electronic interferometers and one takes advantage of the different propagation times in different arms to control the interference pattern. 
The corresponding experiments are still under developments, but first progress has been made in time-resolved processing and sensing of single electron excitations 
 \cite{Roussel21, bartolomei25, souquetbasiege24}
and recent results indicate that the electronic coherence
is in fact robust with respect to high frequency excitations \cite{ouacel2025, Assouline23}.

When the pulse duration is shorter than the electronic time of flight in the different interferometers arms, the short pulse limit, an interesting transient regime should appear where several novel phenomena are predicted: oscillating current with pulse amplitude \cite{gaury14a} or waiting time \cite{gaury14b}, a non-superconducting analogue to the AC Josephson effect \cite{gaury15}, the dynamical control of Majorana-like bond states \cite{weston15} or of RKKY magnetic interactions \cite{meyer17}.  

Accessing this regime with electronic interferometers typically requires voltage pulses of duration as short as a few picoseconds so that the corresponding experiments are difficult and should not be expected for several years.
In this article, we consider the case of a resonant level as can be found in small quantum dots \cite{Kouwenhoven97}. Such a model can be obtained from a simple Fabry-Perot cavity in the limit where the mean level spacing is infinitely larger than the width of the resonance. However, its strength comes from the fact that the model is completely generic and does not requires the fine tuned interference pattern of the Fabry-Perot interferometer. We show that a phenomenon similar to that of the interferometers is present, including the most salient features (see below).
However, the advantage of the quantum dot is that its characteristic frequency scale is easily tunable from kHz to Thz (through a simple electrostatic gate). In particular, its tunneling rate is amenable to a regime in the 10 GHz range, fast enough for the system to still be phase coherent, yet slow enough to be accessible to existing experimental setups.

\begin{figure}[tbh]
	\centering
	\includegraphics[width=85mm]{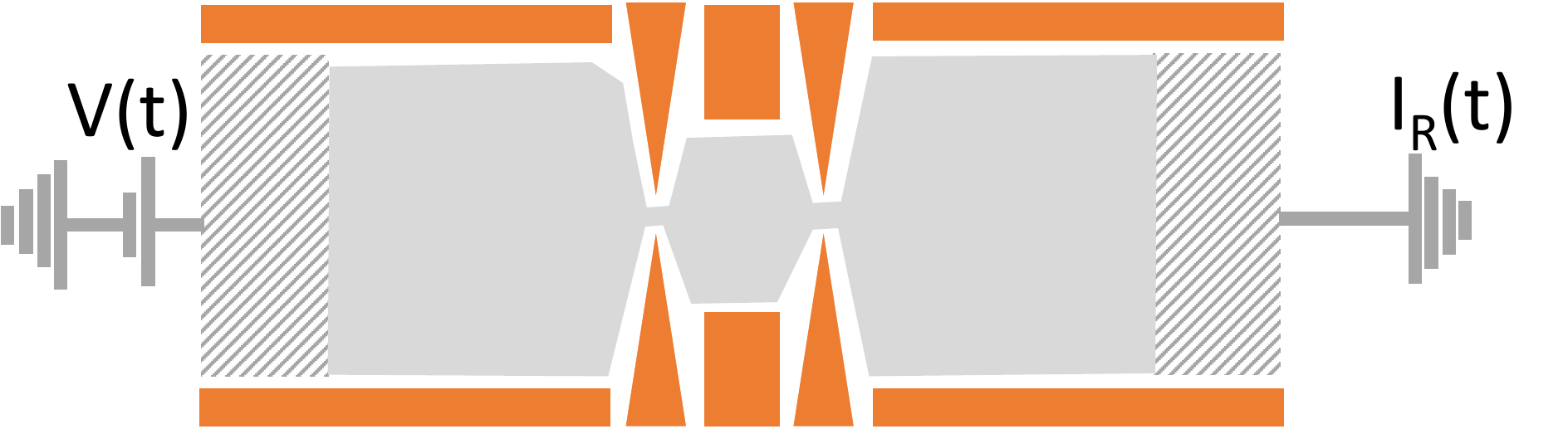}
	\vspace{-1mm}
  \caption{Schematic of the situation considered in this article: a quantum dot defined
  by electrostatic gates (orange surfaces) is connected to two electrodes (dashed surfaces). A voltage pulse $V(t)$ is applied to the left electrode; one measures the current $I_R(t)$ on the right.
}
\label{fig:schematic}
\end{figure}

The specific situation that we consider in this article is very simple 
(see Fig.~\ref{fig:schematic}): a voltage pulse $V(t)$ is applied on an Ohmic contact. This pulse creates  an charge density excitation (hereafter a ``surface plasmon'') which propagates towards the quantum dot. At the dot, part of the pulse is reflected and the remaining part is transmitted through the dot. We study the number of transmitted electron $n_t$ as a function of the number of injected electrons $\bar n$ and the pulse duration $\tau$. In the short pulse limit $\Gamma\tau / \hbar \ll 1$ (the tunneling rate $\Gamma$ is the width of the dot resonance, or equivalently the inverse of the dot level lifetime), we predict that $n_t$
actually \emph{oscillates} with $\bar n$ and can even become \emph{negative}, i.e.\ the current flows against the voltage drop. The origin of these (rather surprising) findings 
can be traced back to the fact that the pulses are not ``classical bullets'' sent in vacuum but rather should be understood as a distortion of the Fermi sea \cite{gaury14a}.

This article begins with a qualitative description of the physics involved, followed by
a discussion of the main result of this article (section \ref{sec:main_idea}). We then continue
with an introduction to the microscopic model used in the simulation (section \ref{sec:model}), together with a brief description of the numerical technique used.
Section \ref{sec:results} describes our results. We study the transmission through the quantum dot as a function of the amplitude and duration of the pulse both analytically and numerically.
Section \ref{sec:conclusion} summarizes the results and discusses how the predicted effects could be
observed experimentally.
The appendices contain the analytical derivations and the details of the numerical procedure.

\section{Understanding particle-wave duality for surface plasmons.}
\label{sec:main_idea}

Particle-wave duality takes a very peculiar form for the surface plasmon generated by a
voltage pulse. We start our discussion with a qualitative description of this duality
as well as its consequences when this plasmon is scattered by a quantum dot.

\subsection{From voltage pulse to plasmons}

Let us first consider what happens \emph{on the left of} the quantum dot when
a voltage pulse is applied to the Ohmic contact. We consider a one-dimensional conductor
and model the pulse by an abrupt drop of the electric potential $V_e(x,t) = V(t)\theta(x_b -x)$ where
$\theta(x)$ is the Heaviside function. The voltage drop occurs at position $x_b$ and we set $x_b = 0$ throughout this paragraph without loss of generality.
This is an accurate model when the system has screening gates close the electron gas, such that the potential drop takes place on a short scale, see the discussion in section 8.4 of \cite{gaury14}. In practice, it is sufficient that the size of the region where the drop happens is shorter than $\hbar v_F/\Gamma$ where $v_F$ is the Fermi velocity.
Before the pulse, the state of the system is composed of simple plane waves 
$\Psi_E(x,t) = e^{ikx-iEt / \hbar}$.
After the pulse, the state has acquired an additional phase $\phi(t)$ and becomes
$\Psi_E(x,t) = e^{ikx-iEt / \hbar -i\phi(t)\theta(-x)}$ with
\begin{equation}
\phi(t) = \frac{e}{\hbar} \int_{-\infty}^{t} dt' V(t')
\end{equation}
We further assume that the dispersion relation may be linearized $E(k) \approx E(k_F) + \hbar v_F (k-k_F)$.
This assumption is correct when both $\hbar /\tau$ and the pulse amplitude are small with respect to the Fermi energy $E_F$, which covers all the situations of interest for the application discussed here.
Then, the wave-function evolves into
\begin{equation}
\Psi_E(x,t) = e^{ikx-iEt / \hbar -i\phi(t-x/v_F)}.
\end{equation} 
In other words, the pulse has created a \emph{phase domain wall} that propagates ballistically in time. This is the wave aspect of the plasmon and it is expected to have
physical consequences if $\phi(t = +\infty)$ is different from a multiple of $2\pi$.

On the other hand, according to the Landauer formula, the pulse
generates an injected current $I_I(t) = e^2/h V(t)$ assuming again a large Fermi energy compared to the pulse characteristic energies (i.e.\ that the pulse is adiabatic with respect to the dispersion relation of the one-dimensional conductor). We find that the number of electrons $\bar n$ injected by the pulse is
\begin{equation}
  \bar{n} = \frac{1}{e} \int_{-\infty}^{\infty} dt I_I(t) =\frac{\phi(t=+\infty)}{2 \pi}.
\label{eq:nbar_def}
\end{equation}
This is the particle aspect of the plasmon: the center of the ``phase domain wall'' carries
a charge. When this charge is not an integer the phase of the plane wave in front of the plasmon is different from the phase behind it.

Things get interesting when this plasmon is injected into a system that
has an internal time scale $\tau_d$ (in this article, $\tau_d=\hbar/\Gamma$). Indeed, in the simplest situation of a system with no internal dynamics, the plasmon is simply transmitted or reflected with a certain probability. Its wave nature becomes only apparent when more complex measurements, such as measuring the quantum fluctuations of the current \cite{levitov96, *levitov97, *keeling06} are performed. 
For example, the current noise displays a sharp minimum when $\bar n$ is an integer. A minimum noise is also found for a particular shape, the Lorentzian, of the pulse. Indeed the Lorentzian corresponds to the least abrupt variaiton of the voltage, hence do not create unnecessary particle-hole excitations. 
If the system \emph{has} an internal dynamic, these kind of features begin to appear in 
common observables such as the current. Let us consider the simple case of a two-path interferometer \cite{gaury14a}: the transmission amplitude $d(E)$ is the sum of two terms
$d(E)= a + b e^{-iE \tau_d / \hbar }$ where the extra phase $e^{-iE \tau_d / \hbar}$ corresponds to the delay between the two paths (for example the difference in lengths if the two paths correspond to two different physical arms). The transmitted wavefunction is now the sum of the two shifted
paths: $\Psi_E(x,t) = e^{ikx-iEt / \hbar}[a e^{-i\phi(t-x/v_F)}+ b e^{-i\phi(t-x/v_F - \tau_d)}]$.
In other words, if the delay $\tau_d$ is large enough compared to the pulse duration, we have caused the front of the pulse to interfere with the back of the pulse. 
This leads to a contribution to the
current that oscillates as $\cos(2\pi\bar n)$, a very non-classical behavior.

The mathematical procedure to transform the above qualitative discussion into a quantitative one has been described in \cite{gaury14, gaury14a} and goes as follows. (i) A Fourier transform is performed on $e^{-i\phi(t)} = \int d\mathcal{E} 
K(\mathcal{E})e^{-i\mathcal{E} t / \hbar} $ to calculate the incident wave-function $\Psi_E(x,\mathcal{E})$, where $E$ is the energy of the electron in the past (before the voltage pulse has been applied) and $\mathcal{E}$ the energy after the pulse has been applied. (ii) The transmitted wave function $\Psi_E(x,\mathcal{E}) d(\mathcal{E})$ is calculated. (iii) Fourier transform the transmitted wave-function back to time to calculate the corresponding current $I(E, t) \propto {\rm Im } \Psi^* \partial_x \Psi$.
(iv) One integrates over all filled states to obtain the actual current $I_R(t) =\int dE \, I(E, t) f(E)$, where $f(E)$ is the Fermi function and $I_R(t)$ is the current flowing to the right of the device. (v) The current is integrated over time 
to give the average number of transmitted electrons
\begin{equation}
  n_t = \frac{1}{e}\int_{-\infty}^{\infty} dt I_R(t).
\label{eq:nt_def}
\end{equation}
$n_t$ and the corresponding effective ``transmission probability'' $n_t/\bar n$ are the main quantities of interest in the present work. In an actual experiment, one would apply a sequence where the pulse is repeated periodically with a period $T$ (typically $1/T$ would be of the order of 1\,MHz or lower, so that this repetition does not affect the results, its only role being to accumulate statistics). In such a setup the measured DC current would be directly proportional to $n_t$: $I_{\rm dc}= en_t/T$. An important aspect is that DC currents are much easier to measure than time-dependent quantities.
 
\subsection{Main results of this work}

\begin{figure}[tbh]
	\centering
	\includegraphics[width=85mm]{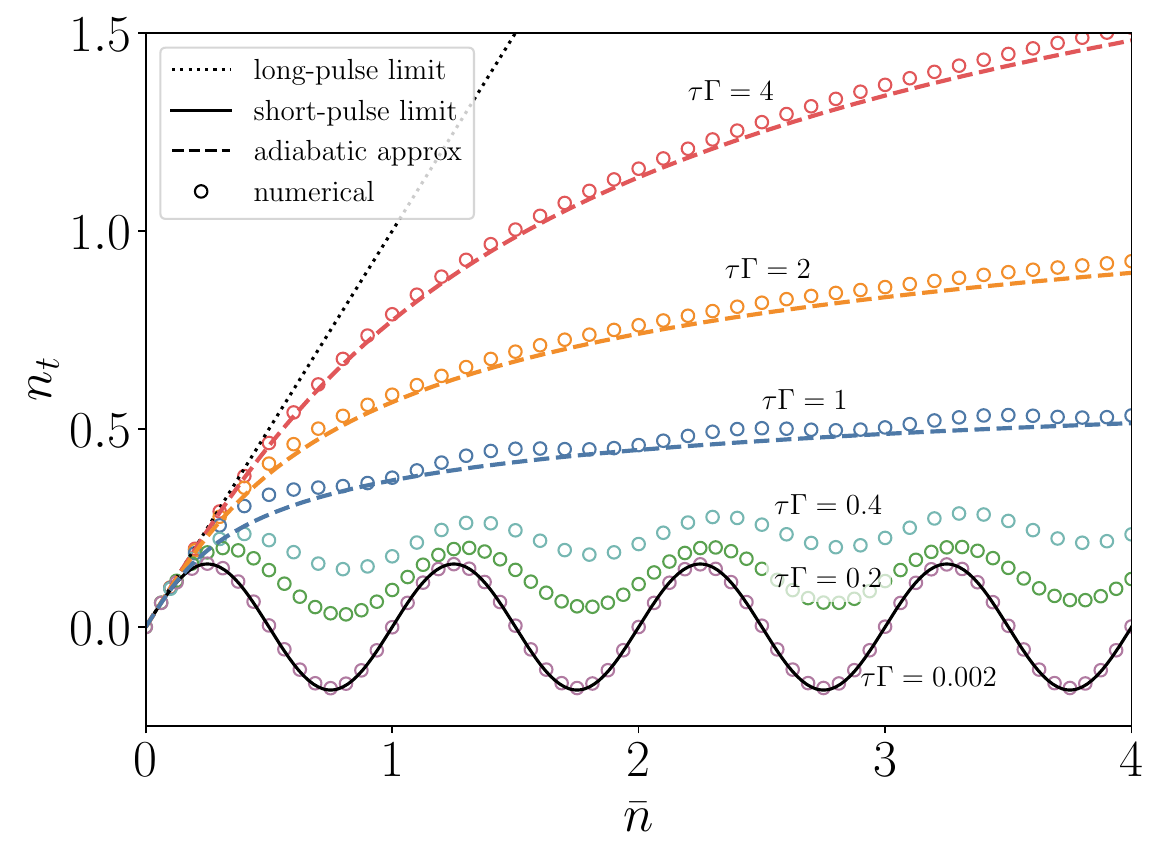}
	\vspace{-1mm}
  \caption{%
Number of transmitted charges $n_t$ through the quantum dot 
as a function of the number of injected charges $\bar{n}$ for various pulse durations $\tau$. 
Short pulse limit $\tau \rightarrow 0$ (solid line, Eq.\ \eqref{eq:shorttime}), long pulse limit $\tau \rightarrow \infty$ (dotted line) and adiabatic approximation (dashed lines, Eq.\ \eqref{eq:adiabatic} for $\tau \Gamma = 4, \, 2$ and $1$). Numerical simulations are performed with Tkwant \cite{Tkwant21} (circles).
In all cases $\epsilon_0 = 0$.
}
\label{fig:slow_fast}
\end{figure}

To simplify the notation, from now on we will measure time in units of $\hbar$, which is equivalent to setting $\hbar = 1$.
The transmission amplitude of a single resonant level is given by a Lorentzian,
\begin{equation}
d(E)= \frac{\Gamma}{(E-\epsilon_0) + i \Gamma}
\label{eq:d_amplitude}
\end{equation}
where $\Gamma$ is the inverse lifetime of the level which is detuned from the Fermi level by $\epsilon_0$ (experimentally, such a detuning is typically controlled directly using a plunger gate).
We consider a Gaussian voltage pulse of duration $\tau$ and amplitude $V_p$ of the form
\begin{align}
  & \qquad V(t) = V_p e^{- t^2/\sigma^2} \nonumber \\
  & \sigma = \tau / (2 \sqrt{\ln 2}) ,  V_p =  4 \sqrt{\pi \ln{2}} \frac{\bar{n}}{\tau}.
\label{eq:gaussian_potential}
\end{align}
For the purpose of this work, the Gaussian pulse is very similar to the Lorentzian one.
However, the Gaussian, having no long tails, is more convenient for numerical simulations.

In the limit of ultrashort pulses $\Gamma\tau\ll 1$, $n_t$ is independent of the pulse shape
and has the following simple form (see Appendix \ref{sec:appendix_short} for the details of the calculation),
\begin{align}
  n_{t} = \frac{1}{2 \pi} \frac{\Gamma}{\epsilon_0^2 + \Gamma^2} \Bigl[ \Gamma \sin(2 \pi \bar{n}) + \epsilon_0 (1 - \cos(2 \pi \bar{n})) \Bigr].
\label{eq:shorttime}
\end{align}
This is the most important result of this work and also the most counter-intuitive limit. In fact, we find that the number of transmitted electrons \emph{oscillates} with the amplitude of the pulse (parametrized by $\bar n$) and can become \emph{negative},
i.e.\ the pulse on the left electrode makes it possible to pump electrons from the right lead.
Although in this regime the electrons flow \emph{against} the potential drop, this does not violate any fundamental principle, since the required energy is provided by the pulse.
The negative current can be understood quantitatively from the dynamical interference phenomena described above. We must first remember that the zero net current observed \emph{at equilibrium} results from the cancellation of the current coming from the left by the current coming from the right.
 What the dynamical interference does is to temporarily block the current coming from the left (destructive interference). During this transient time the current coming from the right is not compensated by the current coming from the left, hence the negative net current. 

In the other extreme (adiabatic) limit $\Gamma\tau\gg 1$, $n_t$ can also be calculated analytically. For a Gaussian shaped pulse in the form of Eq.\ \eqref{eq:gaussian_potential}, the expression is not particularly transparent but it corresponds to a simple monotonic dependence on $\bar n$,
\begin{align}
  n_{t} &= - \frac{\tau \Gamma }{4 \sqrt{\pi \ln 2}} \Im \textrm{Li}_{3/2}\left(\frac{\bar{n} }{\tau}\frac{4 \sqrt{\pi \ln 2}}{\epsilon_0 + i \Gamma} \right ),
\label{eq:adiabatic}
\end{align}
where $\textrm{Li}_s(z)$ is the polylogarithm \cite{AbraSteg72}.
The derivation of above formula is detailed in appendix \ref{sec:appendix_long}.

The entire crossover from $\Gamma\tau\ll 1$ to $\Gamma\tau\gg 1$ is shown in 
Fig.~\ref{fig:slow_fast} which summarizes the main prediction of this work.
The different curves correspond to numerical or analytical approaches, all of which
are in quantitative agreement. The curve passes smoothly from a monotonous
to an oscillatory behavior, with the oscillations becoming visible for $\Gamma\tau\sim 1$.
As we will discuss in Sec.\ \ref{sec:conclusion}, this leaves a relatively comfortable margin for observing these effects with currently available experimental setups.

\section{Microscopic Simulations}
\label{sec:model}

While the two extreme cases of long and ultrashort pulses can be understood analytically,
the calculation of the crossover requires numerical simulations. These simulations are important for predicting the onset of the oscillatory behavior and the negative transmission regime ($n_t<0$). We use Tkwant \cite{Tkwant21, tkwant},
a software developed in the group and which can simulate general time-dependent tight-binding models. Details of the Tkwant simulations are presented in Appendix \ref{sec:appendix_tkwant}.

\subsection{Model}
\begin{figure}[tbh]
	\centering
	\includegraphics[width=85mm]{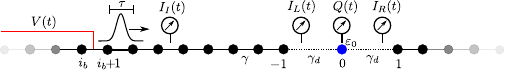}
	\includegraphics[width=40mm]{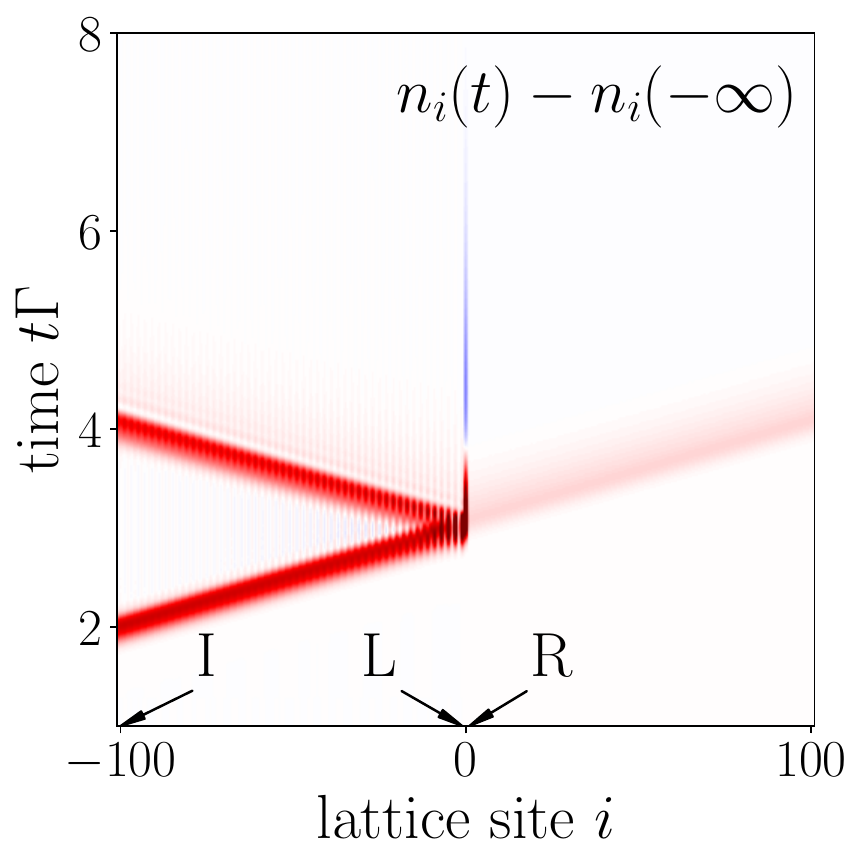}
	\includegraphics[width=42mm]{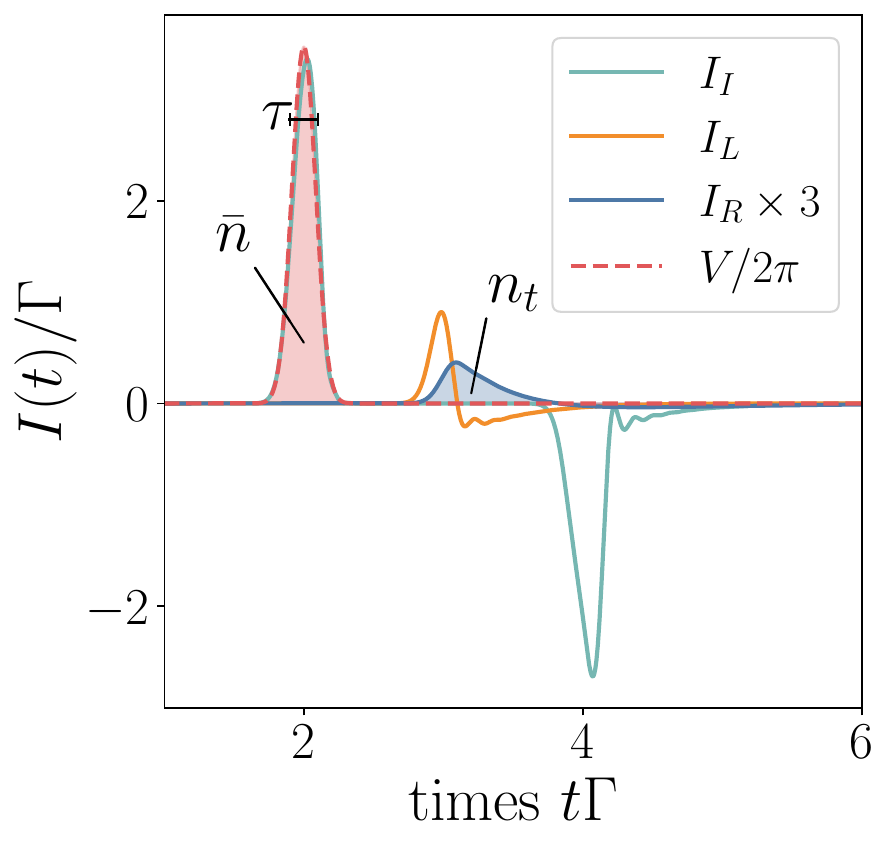}
	\vspace{-1mm}
  \caption{%
Simulations of the microscopic model Eq.\ \eqref{eq:hamiltonian}.
Upper panel: Sketch of the system describing a quantum dot (blue circle, $i=0$) weakly 
connected to two one-dimensional chains on its left and right (black circles, $i\ne 0$),
see text. The current is measured at three positions of interest: injection (I), left of the dot (L) and right of the dot (R).
Lower left panel: color plot of the density $n_i(t) - n_i(-\infty)$ after a pulse injection. Red: positive additional density, blue (at the impurity site): negative extra density.
Lower right panel; current versus time at the three different positions of interest.
Parameters: $\tau \Gamma = 0.2, \bar{n}=0.75, \epsilon_0 = 0$.}
\label{fig:schema}
\end{figure}

The microscopic model used to describe the quantum dot is shown in the upper panel of Figure \ref{fig:schema}. It consists of an infinitely long 1D quantum wire described
by the following tight-binding Hamiltonian,

\begin{equation}
  \hat{H} = \sum_{ij} \gamma_{ij} c^\dagger_i c_j + \sum_i V(t) \theta(i_b - i) c^\dagger_i c_i,
\label{eq:hamiltonian}
\end{equation}
where $c^\dagger_i$ and $c_i$ are fermionic creation and annihilation operators for an electron at discrete lattice site $i$. The off-diagonal part of the first term corresponds to nearest-neighbor hoppings. The coupling is $\gamma_{ij} = - \gamma$ for $|i - j| = 1$, except for the coupling to the impurity site at lattice position 0: 
$\gamma_{01} = \gamma_{-10} = - \gamma_d$ with $|\gamma_d / \gamma| \ll 1$ to form a resonance. The impurity has an additional on-site potential $\gamma_{00} = \epsilon_0$ which accounts for the effects of a plunger gate. Otherwise, $\gamma_{ii}=0$.
A simple Fermi golden rule calculation (exact in this setting) shows that the width
of the resonance is given for this model by,
\begin{equation}
 \Gamma = 2 \gamma^2_d / \gamma.
\label{eq:gamma_def}
\end{equation} 
The second term of the Hamiltonian takes into account the time-dependent bias potential $V(t)$ which is applied uniformly to all sites on the left, such that a voltage drop occurs between site $i_b$ and $i_b +1$ and allows one to inject electrons into the system (see the upper panels of Fig.~\ref{fig:schema}). The current is measured at three different locations: immediately after the injection ($I_I$), to the left of the quantum dot ($I_L$) and to the right ($I_R$); we also measure the charge on the dot $Q$, in particular to verify charge conservation.

A typical output of the simulations is shown in the lower panels of 
Fig.~\ref{fig:schema}.
The lower left panel shows the deviation from the equilibrium electron density
$n_i(t)-n_i(-\infty)$ ($n_i(t) \equiv \langle c^\dagger_i (t) c_i (t)\rangle$) produced by the pulse as a function of the position $i$ and the time $t$. 
One observes the ballistic propagation of the pulse up to the quantum dot where it is
partly reflected and partly transmitted. The lower right panel shows the current
at the three positions of interest and the pulse $V(t)$ itself. The injected current 
$I_I(t)$ (green) precisely matches $V(t)/2\pi$ (dashed red line, using $\hbar = e = 1$ units for convenience), indicating that the simulations are indeed in the regime where the curvature of the dispersion relation can be ignored.
The injected current has an additional negative peak around $t\approx 4/\Gamma$, which corresponds to the current reflected by the dot. The fact that $i_b \ll 0$ allows a clear separation between these two contributions. 
The right current $I_R$ (blue), which is smaller in this example, has been multiplied by 3 to put it on the same scale as the other currents. From these time-dependent currents, the number of injected $\bar n$ and the number of transmitted $n_t$ particles can be obtained by simply integrating the area under the corresponding curves (shaded areas in the figure).

\subsection{Numerical results}
\label{sec:results}

\begin{figure}[tbh]
	\centering
	\includegraphics[width=85mm]{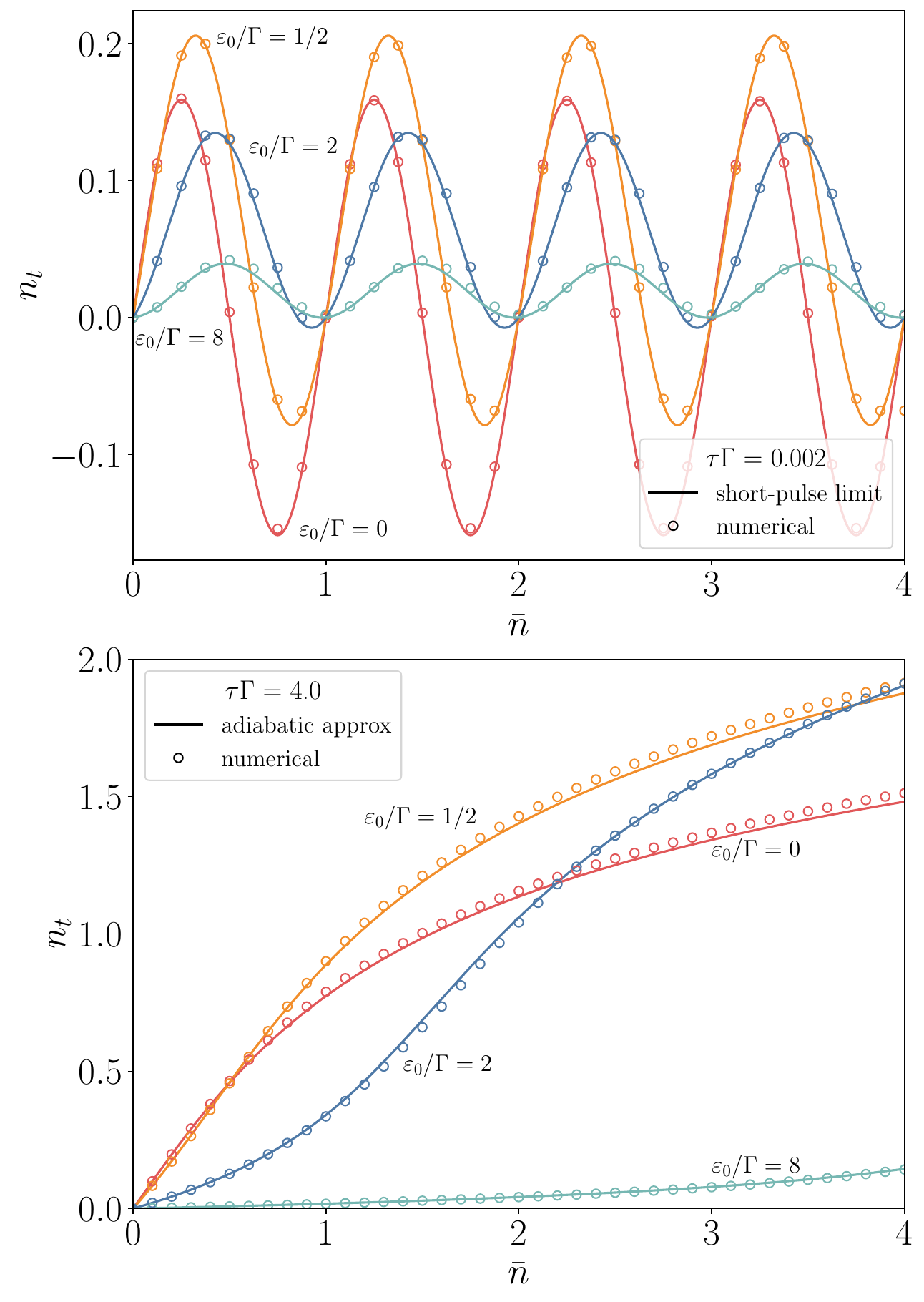}
	\vspace{-1mm}
  \caption{%
Number of transmitted charges $n_t$ as the function of the number of injected charges $\bar{n}$ for different values on the impurity onsite energy $\epsilon_0$ in the short time (upper) and in the long-time limit (lower). Circles are obtained from Tkwant simulations and contineous lines from Eq.\ \eqref{eq:shorttime} in (upper), respectively from from Eq.\ \eqref{eq:adiabatic} in (lower).
  }
\label{fig:nt_nbar}
\end{figure}

The main results of the simulations have already been presented in the discussion of
Figure \ref{fig:slow_fast}. The agreement between the long pulse limit of Eq.\ \eqref{eq:adiabatic} and the numerical data is excellent down to around $\tau \Gamma = 1$.
Around this value an oscillatory behavior becomes visible in the numerical curves. For $\tau \Gamma = 0.002$ the numerical data accurately follow the short pulse theoretical prediction of Eq.\ \eqref{eq:shorttime}.

We present additional data at $\epsilon_0\ne 0$, out of resonance, in Figure \ref{fig:nt_nbar}. The behavior for short and long pulses respectively is shown in the upper and lower panels. Again, one observes a very different behavior in these two limits and a perfect agreement between the theoretical predictions [Eqs.\ \eqref{eq:shorttime} and \eqref{eq:adiabatic}] and the corresponding numerical simulation data.

\begin{figure}[tbh]
	\centering
	\includegraphics[width=85mm]{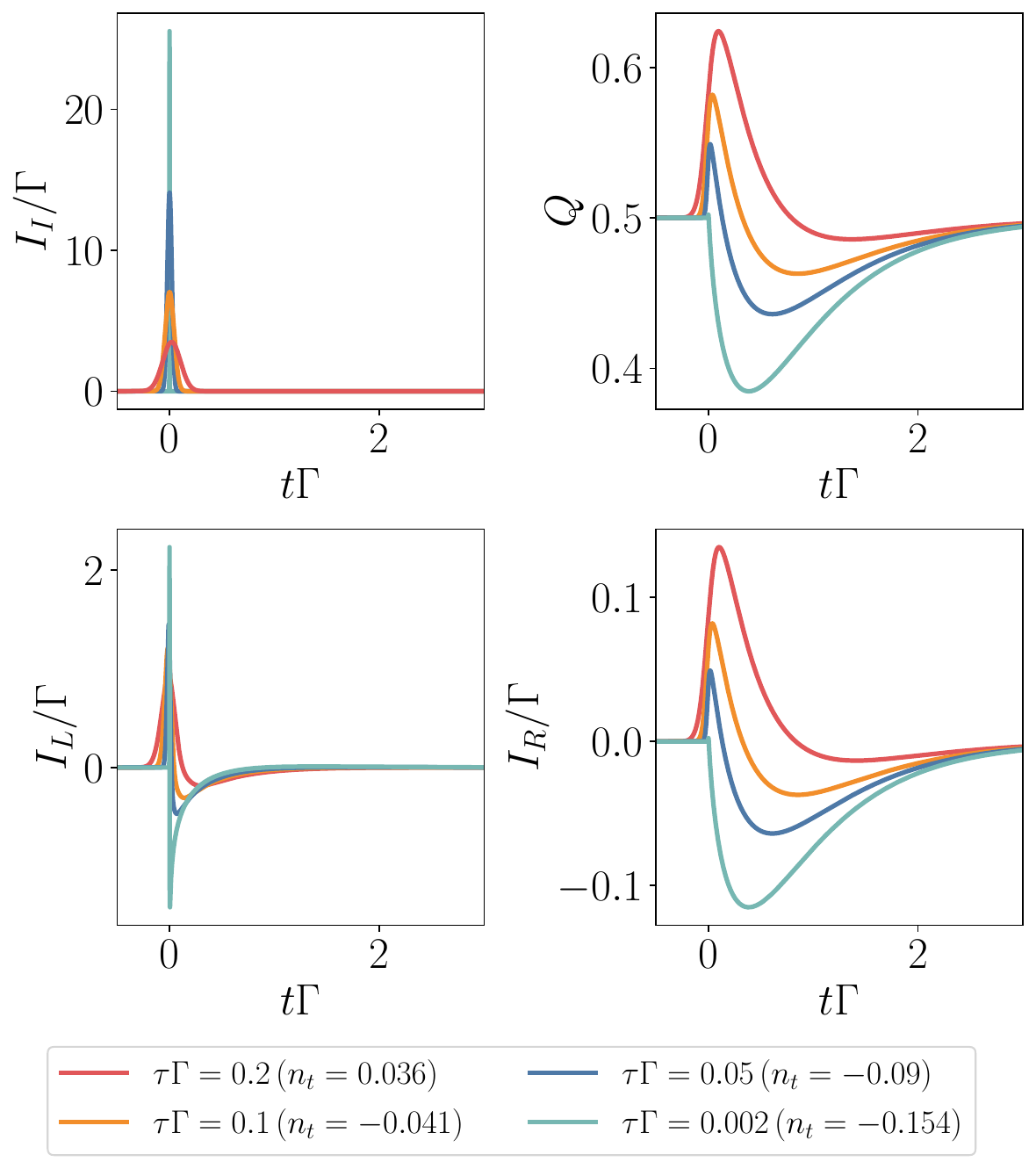}
	\vspace{-1mm}
  \caption{%
Injected current $I_I$ (upper left), onsite charge on the impurity site $Q(t)$ (upper right), left current $I_L(t)$ (lower left), right current $I_R(t)$ (lower right) as a function of time $t$ for different pulse lengths $\tau$. The respective number of transmitted electrons $n_t$ has been obtained according to Eq.\ \eqref{eq:nt_def}. All curves have been shifted horizontally such that the pulse is centered at time $t = 0$ and the $I_I$ curve for the shortest pulse (turquoise) has been multiplied by 0.075 for presentation reasons.
The numerical simulations were performed with Tkwant with parameters: $\bar{n} = 0.75,\, \epsilon_0 = 0$.}
\label{fig:dens_curr_time}
\end{figure}

For completeness, we also show the full time-dependence of the quantum dot charge $Q$ and the currents $I_{L/R}$ in Fig.~\ref{fig:dens_curr_time} for $\bar{n} = 0.75$ (a value for which $n_t<0$ for short pulses). The measurement of e.g.\ $I_R(t)$ is theoretically possible but, as argued above, much more difficult than its integrated value $n_t$. An interesting non-trivial feature is the fact that, depending on the pulse duration $\tau$, the transient evolution of the dot occupation $Q(t)$ shows an increased occupancy (long pulses) or a decreased one (short pulses).

\section{Discussion: experimental relevance of using quantum dots}
\label{sec:conclusion}

The main result of this article is that by repeatedly sending ultrashort pulses
towards a quantum dot, one can measure a DC current that \emph{oscillates} with
the pulse amplitude. This is the onset of a true high-frequency quantum regime.
As discussed in the introduction, such an oscillatory behavior has been predicted before; indeed, its observation is the focus of an intense experimental effort as it is a central  milestone in the construction of an electronic flying qubit \cite{Bauerle18}.

We argue here that the quantum dot is a particularly well suited device to observe this regime as it allows one to relax important experimental constraints. 
Let us discuss the different time scales present in a quantum nanoelectronic device, using GaAs/AlGaAs heterostructures as a reference platform. 
The main timescales are the pulse duration $\tau$, the pulse amplitude $\hbar/V_p$, the characteristic time scale of the device $\tau_d$, the coherence time $\tau_\phi$, the time associated with temperature $\hbar/k_BT$ and the Fermi energy $\hbar/E_F$. 
In electronic interferometers, which are envisioned to build flying qubits, the time scale $\tau_d$ corresponds to the time it takes an electron to travel a length $L$ of one of the interferometer arm. 
With $L$ typically of the order of a few $\mu$m and plasmon velocities of the order of $v_P\approx 10^5$\,m$\cdot$s$^{-1}$, we obtain $\tau_d = L/v_P \approx 10$\,ps. 
It follows that to reach the short time regime $\tau/\tau_d<0.1$, one needs pulses of duration $1$\,ps or less. 
While such short THz pulses may be achievable in the near future, they are currently out of reach, in particular for commercial electronics, especially at dilution refrigerator temperatures.

In a quantum dot, $\tau_d=\hbar/\Gamma$ is fully tunable \emph{in situ} from kHz or lower (used in Coulomb blockade experiments) to the GHz regime when fully open. With a coherence time at $10$\,mK of the order of a few ns, one could target e.g.\  $\Gamma / h \simeq 2$\,GHz for
the device to remain properly coherent. This value also guarantees that $k_B T \ll \Gamma$,
so that thermal broadening is not a limiting factor. The requirement of $\tau/\tau_d = 0.1$ means that $\tau \simeq 50$\,ps, i.e.\ operating at $20$\,GHz. This is well within the capabilities of commercially available arbitrary wave generators. The corresponding voltage amplitude for generating a single electron $\bar n = 1$ is of the order of $V_p \simeq 10\,\mu$V,
for which there is some preliminary experimental evidence that the generated plasmon remains coherent \cite{ouacel2025}. Real quantum dots have two additional energy scales:
the mean level spacing $\delta$ between levels and the charging energy $E_C$. The current
study corresponds to the limit of infinite $\delta$ (a single level is considered). In
a preceding publication we studied the limit of very small $\delta$ (Fabry-Perot regime)
\cite{gaury14a} and also observed an oscillatory ultrafast regime. Therefore, we conclude that while a finite value of $\delta$ is likely to affect our results quantitatively, but qualitatively they would hold. Furthermore, a relatively large value of
$E_C$ would prevent other levels from being populated by the pulse, hence most likely stabilizing the limit studied in this article.

An important but difficult aspect that we have not studied in this article is the role of electron-electron interactions. It was recognized very early, e.g.\ by B\"uttiker \cite{Buttiker93}, that electron-electron interactions must be treated at some level when dealing with time-dependent transport, since otherwise some basic properties such as current conservation can be violated (by disregarding displacement currents). However, the present calculations are explicitly ``gauge invariant'' as shown in \cite{gaury14}. Nevertheless, strictly speaking, the present results are only valid in the presence of a very close metallic gate that screens the Coulomb interaction (in particular its long-range part).
Such gates have actually been recently introduced in GaAs heterostructures \cite{Manfra20} and are already quite common in graphene through graphene-BN-graphite stacks \cite{ronen21}.

Interaction is known to play a role on several levels and has been studied in the context of transport through quantum dots e.g.\ Refs.\ \cite{Fujisawa06, Uimonen11, Pertsova13, Vovchenko_14}. First, it renormalizes the Fermi velocity into the plasmon velocity, which is typically much faster by a factor ten, possibly more. Second, interaction is responsible for the main decoherence channel through the two-particle-one-hole excitations \cite{Altshuler82}. While decoherence has been well studied in DC both experimentally and theoretically \cite{Marguerite16, Assouline23, jo22}, very little, however, is known about it in presence of pulses. The recent results in \cite{Assouline23,ouacel2025} indicates a rather strong resilience of coherence, despite the rather high voltages being used.
Third, electron-electron interactions can be the origin of some genuine correlated physics
such as the Kondo effects, and quantum dots hold an interesting place in this regards. As it is a relatively simple system, there is a number of techniques to study it theoretically in a controlled way and a full solution of the propagation of a voltage pulse through a quantum dot might be within reach. In particular, the recent approach of \cite{Fernandez22}, which systematically calculates the effect of high order Feynman diagrams, is directly formulated for out-of-equilibrium calculations and extensible to pulse dynamics.

We leave the corresponding analysis, which is far beyond the scope of this paper, to a later work.
 Our preliminary calculations performed at the random phase approximation
level (which is available within Tkwant \cite{kloss18} and fully accounts for displacement currents and plasmon velocity renormalization) indicate that our main conclusions, in particular the oscillations in the short pulse limit and the associated negative current, are robust to the presence of electron-electron interactions.

\section*{ACKNOWLEDGMENTS}

T.\,K.\ and X.\,W.\ acknowledge funding from the European Union’s Horizon 2020 research and innovation program under Grant agreement No. 862683 (UltraFastNano), from the French ANR DADDI and T-KONDO and from State aid managed by the Agence Nationale de la Recherche under the France 2030 program, reference ANR-22-PETQ-0012 (EQUBITFLY).
T.\,K.\ greatly appreciates the hospitality of C.\ Bauerle in his group. We thank him as well as S.\ Ouacel, L.\ Mazzella, M.\ Aluffi and T.\ Vasselon for interesting discussions.

\section*{DATA AVAILABILITY STATEMENT}
Python scripts to perform the numerical simulations, the generated data from these simulations, and the scripts to generate the plots in this article are openly available \cite{supplementary_data}.
The Tkwant code is free software and openly available \cite{tkwant}.

\begin{appendix}

\section{Derivation of $n_t(\bar n)$  in the short-pulse limit}
\label{sec:appendix_short}
\renewcommand{\theequation}{A\arabic{equation}}
\setcounter{equation}{0}
This appendix contains the derivation of Eq.\eqref{eq:shorttime}. 
It follows closely a similar derivation down in \cite{gaury14a} using the formalism of \cite{gaury14}. Our starting point is equation (95) from reference \onlinecite{gaury14} which we rewrite  as follows
\begin{equation}
  n_t = \iint_{-\infty}^{\infty} \frac{dE dE'}{(2 \pi)^2} |K(E)|^2 |d(E')|^2 [f(E' - E) - f(E')],
  \label{eq:nt_scatt}
\end{equation}
where the probability amplitude $d(E,E')$ has been decomposed as $d(E,E') = K(E-E') d(E)$.
$K(E-E')$ is an inelastic contribution originating from the voltage drop and
is given by the Fourier transform of
$\phi(t)$ defined in Eq.\ \eqref{eq:phi2}:
\begin{equation}
  K(E) = \int_{-\infty}^{\infty} dt e^{i \phi(t)} e^{- i E t}.
  \label{eq:k_def}
\end{equation}
The second contribution $d(E)$ is the transmission amplitude trough the impurity, which in the weak-coupling regime is given by the Lorentzian in Eq.\ \eqref{eq:d_amplitude}. The corresponding transmission probability has Breit-Wigner form
\begin{equation}
\label{eq:breit_wigner}
  D(E) = |d(E)|^2 = \frac{\Gamma^2}{(E - \epsilon_0)^2 + \Gamma^2}.
\end{equation}
For later reference we also give the Fourier transformed Breit-Wigner formula
\begin{equation}
\label{eq:breit_wigner_fourier}
  \int_{-\infty}^{\infty} dt e^{i \omega t} e^{- \Gamma |t|} = \frac{2 \Gamma}{\omega^2 + \Gamma^2}.
\end{equation}
At zero temperature and for $E_F = 0$, the Fermi function is $f(E) = \theta(-E)$ and Eqs.\ \eqref{eq:nt_scatt}, \eqref{eq:k_def} and \eqref{eq:breit_wigner_fourier} can be combined to
\begin{align}
  n_t &= \frac{\Gamma}{2 (2 \pi)^2} \iint_{-\infty}^{\infty}  dE dE' \iint_{-\infty}^{\infty}  dt dt' e^{i \phi(t) - i \phi(t') - i E (t - t')} \nonumber \\
  & \quad \times \int_{-\infty}^{\infty} du 
    e^{i (E' - \epsilon_0) u - \Gamma |u|} [\theta(E - E') - \theta(-E')] \nonumber \\
&= \frac{i \Gamma}{2 (2 \pi)^2} \iiint_{-\infty}^{\infty} dE  dt dt' e^{i \phi(t) - i \phi(t') - i E (t - t')} \nonumber \\
  & \quad \times\int_{-\infty}^{\infty} du  e^{- i \epsilon_0 u - \Gamma |u|} \frac{(1 - e^{i E u})}{u}     
    \label{eq:nt_scatt1}
\end{align}
where we have used that
\begin{align*}
 \int_{-\infty}^{\infty} d E' e^{i E' u} [\theta(E - E') - \theta(-E')] = \frac{i}{u} (1 - e^{i E u}).
\end{align*}
Moreover, as
\begin{align*}
 \int_{-\infty}^{\infty} d E e^{- i E (t - t')} (1 - e^{i E u}) = 2 \pi [\delta(t - t') - \delta(t - t' - u)],
\end{align*}
equation \eqref{eq:nt_scatt1} can be further rewritten as
\begin{align}
  n_t &= \frac{i \Gamma}{4 \pi} \iiint_{-\infty}^{\infty} du dt dt' \frac{1}{u} e^{i \phi(t) - i \phi(t') }  e^{- i \epsilon_0 u - \Gamma |u|} \nonumber \\
  & \quad \times [\delta(t - t') - \delta(t - t' - u)] \nonumber \\
      &= \frac{i \Gamma}{4 \pi} \iint_{-\infty}^{\infty} du dt \frac{1}{u} \left( 1 - e^{i \phi(t) - i \phi(t - u) } \right)  e^{- i \epsilon_0 u - \Gamma |u|} .
    \label{eq:nt_scatt2}
\end{align}
In the short-pulse limit, $\phi(t) = 2 \pi \bar{n} \theta(t)$, such that
\begin{align*}
 \int_{-\infty}^{\infty} d t \left( 1 - e^{i \phi(t) - i \phi(t - u) } \right) = u \, \sgn(u)  \left( 1 - e^{2 \pi i \bar{n} \sgn(u)} \right),
\end{align*}
from which we finally obtain the result as
\begin{align}
  n_t &= \frac{i \Gamma}{4 \pi} \int_{-\infty}^{\infty} du   e^{- i \epsilon_0 u - \Gamma |u|} \sgn(u)  \left( 1 - e^{2 \pi i \bar{n} \sgn(u)} \right) \nonumber \\
&= \frac{1}{2 \pi} \frac{\Gamma}{\epsilon_0^2 + \Gamma^2} \Bigl[ \Gamma \sin(2 \pi \bar{n}) + \epsilon_0 (1 - \cos(2 \pi \bar{n})) \Bigr].
\end{align}

\section{Derivation of $n_t(\bar n)$ in the (adiabatic) long-pulse limit}
\label{sec:appendix_long}
\renewcommand{\theequation}{B\arabic{equation}}
\setcounter{equation}{0}
In this section, we provide the derivation of Eq.\ \eqref{eq:adiabatic}.
In DC, the current $I$ trough the wire can be calculated from the standard Landauer formula.
\begin{align}
 I(V) =  \int_{E_F}^{E_F + V} \frac{d E}{2 \pi} D(E).
\end{align}
In the adiabatic limit, where the pulse in infinitely slow ($\tau \gg 1/\Gamma$),
the Landauer formula trivially holds by replacing $V$ (resp. $I$) by $V(t)$ (resp. $I(t)$), 
\begin{align}
 I(V(t)) =  \int_{E_F}^{E_F + V(t)} \frac{d E}{2 \pi} D(E).
\end{align}
Writing this formula in our special setup, with $E_F = 0$ and describing the transmission trough the wire by Breit-Wigner formula Eq.\ \eqref{eq:breit_wigner}, one finds
\begin{align}
 I(V(t)) &= \int_{0}^{V(t)} \frac{d E}{2 \pi} \frac{\Gamma^2}{(E - \epsilon_0)^2 + \Gamma^2} \nonumber \\
   &= \frac{\Gamma}{2 \pi} \left( \arctan\left(\frac{V(t) - \epsilon_0}{\Gamma} \right) + \arctan\left(\frac{\epsilon_0}{\Gamma} \right) \right).
   \label{eq:adiabat1}
\end{align}
The number of transmitted charges $n_t$ is obtained according to Eq.\ \eqref{eq:nt_def} by integrating the above equation for the current over the time. 

Before continuing, let us show that the relation Eq.\ \eqref{eq:adiabat1} can be derived as well by taking the adiabatic limit of the ``pulse conductance matrix'' formalism of appendix \ref{sec:appendix_short}.
For this, we start with the expression for $n_t$ from  Eq.\ \eqref{eq:nt_scatt2} (established \emph{before} taking the short pulse limit):
\begin{align}
  n_t = \frac{i \Gamma}{4 \pi} \int_{-\infty}^{\infty} du  \frac{1}{u} e^{- i \epsilon_0 u - \Gamma |u|} \int_{-\infty}^{\infty} dt \left( 1 - e^{i \phi(t) - i \phi(t - u) } \right).
\label{eq:nt_scatt3}
\end{align}
Due to the $e^{- \Gamma |u|} / u$ factor, only small values in $u$ with $|u| \ll 1/\Gamma$ will effectively contribute to the integral. Hence, in the adiabatic limit where $\phi$ varies slowly with respect to its argument, it can be developed as
\begin{align}
 \phi(t - u) \simeq \phi(t) -  V(t) u.
\end{align}
From this, Eq.\ \eqref{eq:nt_scatt3} becomes
\begin{align}
  n_t &\simeq \frac{i \Gamma}{4 \pi} \int_{-\infty}^{\infty} du  \frac{1}{u} e^{- i \epsilon_0 u - \Gamma |u|} \int_{-\infty}^{\infty} dt \left( 1 - e^{i V(t) u } \right)
\nonumber \\ 
&= \frac{\Gamma}{2 \pi} \int_{-\infty}^{\infty} dt  \int_{0}^{\infty} du \frac{ e^{-\Gamma u}}{u} \left( \sin(\epsilon_0 u) - \sin([\epsilon_0 - V(t)] u)  \right).
\end{align}
Using  the identity
\begin{equation}
	\int_{0}^{\infty} du \frac{e^{- \Gamma u}}{u} \sin(\epsilon_0 u) = \arctan(\epsilon_0 / \Gamma), \quad \Gamma > 0
\end{equation}
one can write above equation as
\begin{align}
  n_t =  \frac{\Gamma}{2 \pi} \int_{-\infty}^{\infty} dt \left( \arctan\left(\frac{V(t) - \epsilon_0}{\Gamma} \right) + \arctan\left(\frac{\epsilon_0}{\Gamma} \right) \right).
\label{eq:adiabat2}
\end{align}
The above equation \eqref{eq:adiabat2} is strictly identical to the integral over time of
Eq.\ \eqref{eq:adiabat1} obtained by a simpler argument.

To calculate the remaining integral over the time, we specify the pulse to be of Gaussian form given in Eq.\ \eqref{eq:gaussian_potential}.
Using partial integration and $d/ dx [\arctan{x}]  = 1 / (1 + x^2)$ one can rewrite Eq.\ \eqref{eq:adiabat2} as
\begin{equation}
	n_t = \frac{\Gamma^2}{\pi \sigma^2} \int_{-\infty}^\infty dt \, t^2 \frac{V(t)}{(V(t) - \epsilon_0)^2 + \Gamma^2}.
\end{equation}
Moreover, after factorizing the integrand and with the help of
\begin{equation}
	\int_{-\infty}^\infty dx  \frac{x^2 e^{-x^2}}{a e^{-x^2} +  b - i} = - \frac{\sqrt{\pi}}{2 a} \textrm{Li}_{3/2} \left(\frac{a}{i - b}\right),
\end{equation}
where
\begin{equation}
	\textrm{Li}_s(z) = \sum_{k=0}^\infty \frac{z^k}{k^s}
\end{equation}
is the so-called Polylogarithm \cite{AbraSteg72},
we finally obtain
\begin{equation}
	n_t = -\frac{\Gamma \sigma}{2 \sqrt{\pi}} \Im \textrm{Li}_{3/2}\left( \frac{V_p }{\epsilon_0 + i \Gamma}\right).
\end{equation}
Substituting the definitions of the Gaussian pulse from Eq.\ \eqref{eq:gaussian_potential} into above formula, we arrive at Eq.\ \eqref{eq:adiabatic}.

It is helpful for a qualitative understanding to develop $\textrm{Li}_{3/2}(z)$ for large and small arguments $z$. For $|z| \ll 1$, $\textrm{Li}_{3/2}(z) = z + \mathcal{O}(z^2)$, such that Eq.\ \eqref{eq:adiabatic} becomes
\begin{equation}
	n_t \simeq \bar{n} \frac{\Gamma^2}{\epsilon_0^2 + \Gamma^2}.
\end{equation}
The linear dependence $n_t \sim \bar{n}$ can be observed also in Figure \ref{fig:slow_fast} and in the lower panel of Figure \ref{fig:nt_nbar}.
For large argument $|z| \gg 1$, $\textrm{Li}_{3/2}(z) = - 2 \sqrt{-\pi \ln{z}} + \mathcal{O}(1/z^2)$.
For $\epsilon_0 = 0$ we find the scaling behavior for large $V_p / \Gamma$ or equivalently large $\frac{\bar{n}}{\tau \Gamma}$ as
\begin{align}
	n_t &\simeq \sigma \Gamma  \sqrt{\ln{(V_p / \Gamma)}} 
	\sim \tau \Gamma \sqrt{\ln{\left(\frac{\bar{n}}{\tau \Gamma} \right)}}.
\end{align}
Unfortunately this assymptotic formula is only quantitative for very large pulse amplitudes
and in practice does not fit the numerical data quantitatively; the full form Eq.\ \eqref{eq:adiabatic} must be used instead.

\section{Details of the numerical simulations}
\label{sec:appendix_tkwant}
\renewcommand{\theequation}{C\arabic{equation}}
\setcounter{equation}{0}

This appendix describes the main formalism used in the numerics as well as
additional numerical data. The method is based on a wavefunction formalism described in \cite{gaury14, Tkwant21}. 
The calculations for this article have been performed using \textsc{Tkwant} v.\ $1.1.0$, \textsc{kwant} v.\ $1.4.1$ and \textsc{kwantSpectrum} v.\ $0.1.1$.

\subsection{Notations and basic formalism}
The first step in the calculation is to use a gauge transformation as described, for example, in Ref.\ \cite{Tkwant21}. This amounts to rewriting the original Hamiltonian in Eq.\ \eqref{eq:hamiltonian} into a form where the explicit time dependence appears only on the coupling element between the two sites $i_b$ and $i_{b}+1$ where the voltage drop occurs. Hence, the explicit time dependence is only present in the central finite region, not in the 
infinite electrodes, which facilitates the calculations \cite{gaury14}. The phase is defined (in units of $e / \hbar$) as
\begin{align}
  \phi(t) = \int_{- \infty}^t dt'\  V(t') ,
\label{eq:phi2}
\end{align}
and we obtain for the Gaussian potential in Eq.\ \eqref{eq:gaussian_potential} 
\begin{align}
  \phi(t) = A (1 + \textrm{erf}( t / \sigma)), \qquad A = V_p \sigma \sqrt{\pi}/2.
\label{eq:phi_gauss}
\end{align}
The Hamiltonian in Eq.\ \eqref{eq:hamiltonian} can  be cast in the form
\begin{align}
  \hat{H}(t) = & - \sum_{i} \gamma_{i} c^\dagger_{i + 1} c_i  - \gamma [e^{- i \phi(t)} - 1] c^\dagger_{i_b + 1} c_{i_b} + \textrm{h.c.}  \nonumber \\
  & + \epsilon_0 c^\dagger_{0} c_{0}
\label{eq:h_eff}
\end{align}
where $\gamma_{i} = \gamma$ for all sites $i$, except for the coupling between the dot and the leads, which is $\gamma_{0} = \gamma_{-1} = \gamma_d$.
Before the pulse, the system is initially at equilibrium. We assume that the pulse starts after a time $-t_0$ (in practice $t_0$ is equal to several times the pulse width $\tau$).
The method follows the time evolution of individual one-body wave functions $\psi_{\alpha E}(t, j)$ by solving the time-dependent Schrödinger equation
\begin{subequations}
\label{eq:Tkwant23}
\begin{align}
\label{eq:Tkwant2}
i \partial_t \psi_{\alpha E}(t, j) = \sum_k\mathbf{H}_{jk}(t) \psi_{\alpha E}(t,k) ,
 \\
\psi_{\alpha E}(t<-t_0,j) = \psi_{\alpha E}(j) e^{-i E t}
\label{eq:Tkwant3}
\end{align}
\end{subequations}
where $\mathbf{H}$ is the Hamiltonian matrix, whose matrix elements are defined in terms of the above Hamiltonian as $\hat{H}(t) = \sum_{ij} \mathbf{H}_{ij}(t) c^\dagger_{i} c_j$, $E$ is the energy of the incoming wave, $\alpha \in \{L, R\}$ is the lead index and
$i$ accounts for a lattice site. The scattering states $\psi_{\alpha E}(i)$ are eigenstates of the Hamiltonian for $t<-t_0$. They are a
direct output of Kwant, see Ref.\ \cite{Waintal24} for their precise definitions.

The time-dependent electron density at site $i$ is $n_i(t)$ and $I_{ij}(t) = - I_{ji}(t)$ is the current from site $j$ to site $i$.
Both quantities can be calculated explicitly from the above wave function approach, see Refs.\ \cite{gaury14, Tkwant21}.
The electron density is obtained by summing over all leads $\alpha$ and integrating over all one-body contributions of the initially occupied states.
In units of elementary charges $e$, the electron density is
\begin{equation}
n_i(t) \equiv \langle  c^\dagger_i(t)  c_i(t) \rangle
= \sum_{\alpha} \int \frac{dE}{2 \pi} f_\alpha(E)  |\psi_{\alpha E}(t,i)|^2,
\label{eq:n_tkwant}
\end{equation}
where $f_\alpha$ is the Fermi function for the lead $\alpha$. At zero temperature, 
$f_{\alpha}(E) = \theta(E_F - E)$, where $E_F$ is the Fermi energy and $\theta$ is the Heaviside step function.
The current (in units of $e \gamma / \hbar$) is calculated as
\begin{align}
   I_{jk}(t) \equiv i [\langle  c^\dagger_k(t) \mathbf{H}_{kj}(t)  c_j(t) \rangle
- \langle  c^\dagger_j(t) \mathbf{H}_{jk}(t)  c_k(t) \rangle ] \nonumber \\
 = 2  \Im \int \frac{d E}{2 \pi} \sum_{\alpha} f_\alpha(E) \psi_{\alpha E}^*(t, j)  H_{jk}   \psi_{\alpha E}(t, k).
\label{eq:i_tkwant}
\end{align}

Density and current fulfill the general continuity relation
\begin{equation}
  \partial_t n_{i}(t) = \sum_{j} I_{ij}(t).
\end{equation}
In this manuscript we use the convention that the impurity is located at lattice site $0$. For convenience we also define $Q(t) = n_0(t)$ and $I_L(t) = I_{0,-1}(t)$ and $I_R(t) = I_{1, 0}(t)$, such that a positive sign in $I_{L/R}$ corresponds to a current flowing from left to right. The continuity relation at the impurity thus becomes
\begin{equation}
  \partial_t Q(t) = I_L(t) - I_R(t).
\label{eq:continuity}
\end{equation}

In practice, the numerical solution of the time-dependent Schrödinger equation Eq.\ \eqref{eq:Tkwant23} and of the observables Eqs.\ \eqref{eq:n_tkwant} and \eqref{eq:i_tkwant} are computed with the help of the Tkwant \cite{Tkwant21} package.
To calculate the number of transmitted charges $n_t$ from Eq.\ \eqref{eq:nt_def},
the current $I_R(t_i)$ is computed on a set of timepoints $t_i$.
The data is then interpolated using cubic spline interpolation and finally integrated numerically. We provide example Python scripts as supplementary material which show the details of our procedure.

\subsection{Additional numerical data}

This section comprises additional analyses and verifications of the system and methods.

\subsubsection{Static DC analysis of the wire and impurity}
\label{sec:appendix_dc}

\begin{figure}[tbh]
	\centering
	\includegraphics[width=90mm]{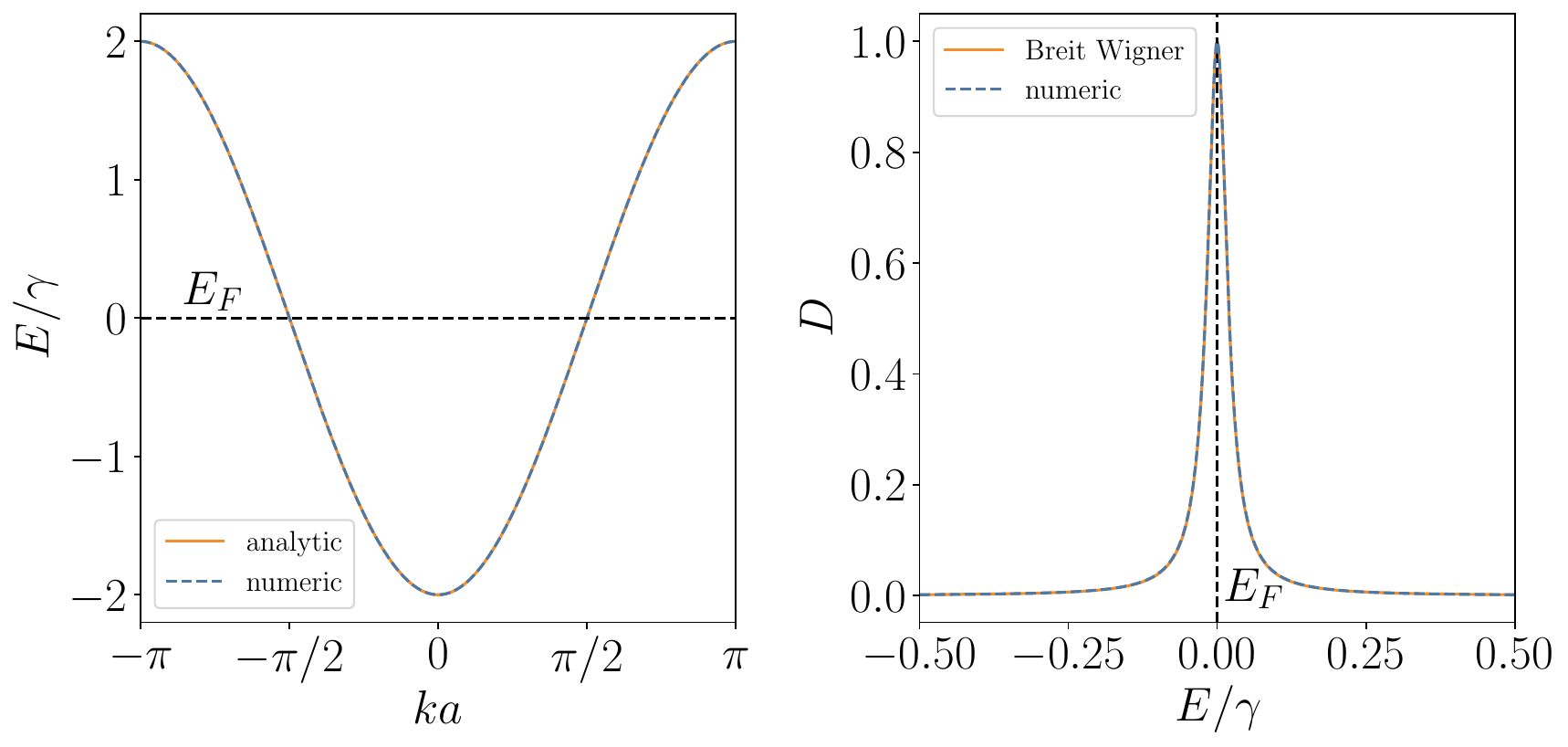}
	\vspace{-1mm}
	\caption{
	Static DC properties of the 1D quantum wire and the impurity.
	Left panel: Energy dispersion $E(k)$ as a function of the momentum $k$ for the infinite long tight-binding chain with nearest-neighbor coupling $\gamma$ and gridspacing $a$. Exact analytical expression $E(k) = - 2 \gamma \cos(k a)$ (straight yellow) and numerical result obtained from Kwant \cite{groth14} (blue dashed).
The black dotted line corresponds to the Fermi energy $E_F = 0$, which is used throughout the article. Right panel: Transmission $D(E)$ of the impurity as a function of the energy $E$ of the incoming wave. Analytic Breit-Wigner transmission from formula Eq.\ \eqref{eq:breit_wigner} (straight yellow), numerical result from the Kwant software \cite{groth14} (blue dashed). Parameters: $|\gamma_d / \gamma| = 0.1, \epsilon_0 = 0$.
	}
\label{fig:dc}
\end{figure}

Prior to any time-dependent simulation, a static DC analysis must be performed to determine the parameters of the system. The left panel of Figure \ref{fig:dc} shows the dispersion energy $E(k)$ as a function of the momentum $k$ in the first Brillouin zone for each of the two semi-infinite leads of the Hamiltonian Eq.\ \eqref{eq:hamiltonian}. $E(k)$ has been calculated using the Kwant software \cite{groth14} and is identical to the cosine dispersion that can be derived analytically for the discrete tight-binding chain.
The black dashed line indicates the Fermi energy $E_F$, which we set to zero throughout this article. The right panel of Figure \ref{fig:dc} shows the transmission $D(E)$ of the impurity as a function of energy. The transmission is sharply peaked and has a Lorentzian shape similar to the Breit-Wigner transmission Eq.\ \eqref{eq:breit_wigner}. The Fermi energy  $E_F = 0$ corresponds to the center of the curve where $D(E_F) = 1$.

\subsubsection{Effect of finite bandwidth in the simulations}
\label{sec:appendix_wire}

In this section we estimate the smallest pulse duration that can be used so that the simulations \emph{are not} affected by the microscopic description of the device.
In fact, the discrete tight-binding Hamiltonian Eq.\ \eqref{eq:hamiltonian} has two different energy scales: First, a low-energy scale $\Gamma$ related to the impurity and second a high-energy scale $\gamma$ resulting from the discreteness of the lattice.
We are interested in a regime where the results are essentially independent of $\gamma$
since our microscopic description is not particularly realistic. More precisely, we are looking for
an intermediate regime where
\begin{equation}
	\gamma^{-1} \ll \tau \ll \Gamma^{-1}
\end{equation}
holds. The first inequality guarantees that the microscopic description plays a role only through its property at the Fermi level, while the second is the ultrashort pulse limit. To discuss the first inequality, it is sufficient to consider a transparent 1D wire without impurities (Eq.\ \eqref{eq:hamiltonian} with $\gamma_d = \gamma$ and $\epsilon_0 = 0$). The definition for $n_t$ in Eq.\ \eqref{eq:nt_def} thus corresponds to the effective number of particles injected into the system.

\begin{figure}[tbh]
	\centering
	\includegraphics[width=60mm]{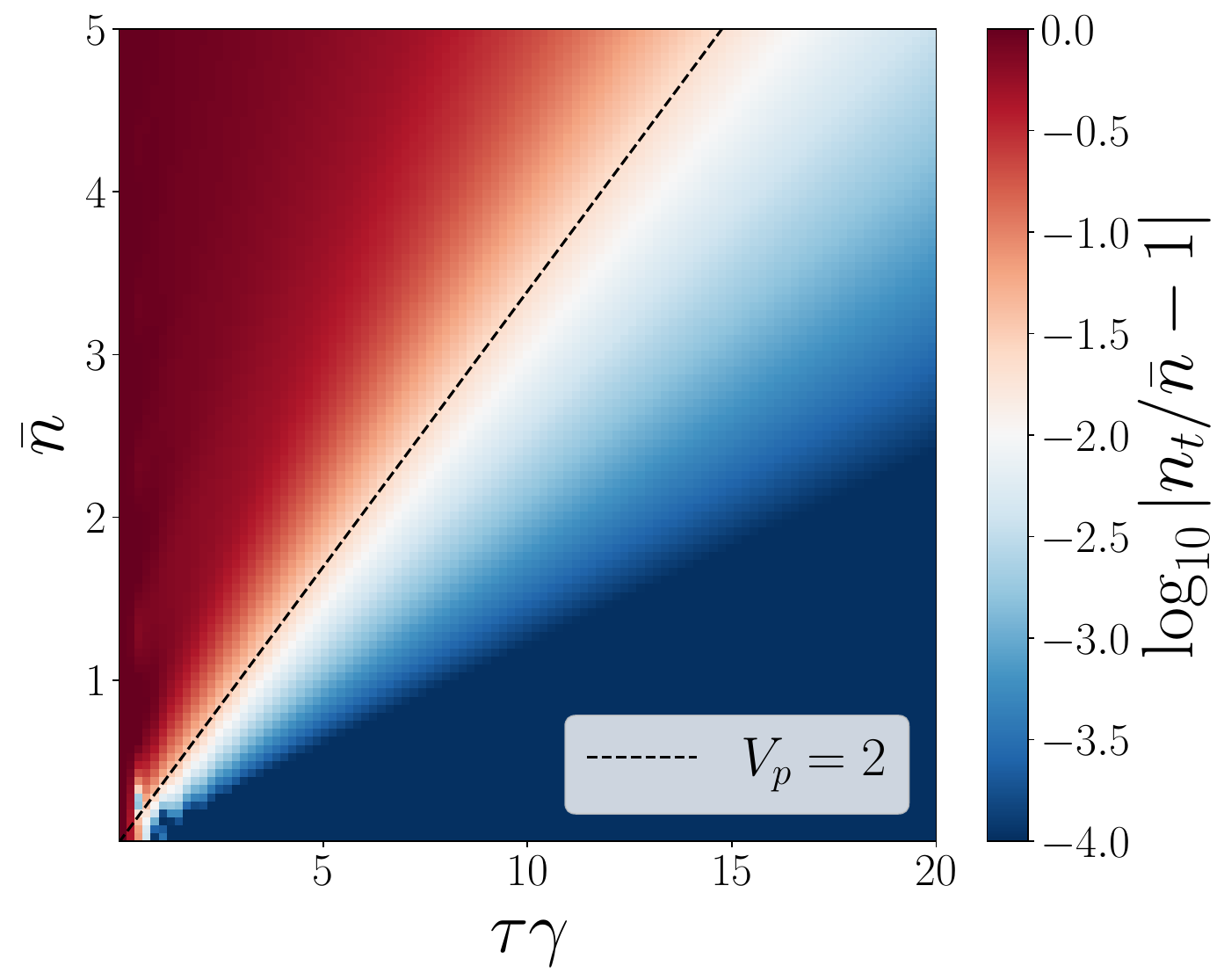}
	\vspace{-1mm}
	\caption{Relative mismatch between the effective number of injected electrons $n_t$ and the theoretical number of injected electrons $\bar{n}$ for a transparent 1D tight-binding wire. The mismatch is plotted as a function of the Gaussian pulse parameters $\bar{n}$ and $\tau$ Eq.\ \eqref{eq:gaussian_potential}.
The black vertical line marks the point where the maximum of $V(t)$ reaches the upper bandgap such that $V_p = 2$ and limitations due to the bandgap will additionally contribute in the upper left triangle above the line.
}
\label{fig:nt_inject}
\end{figure}

When the pulse length becomes too short, such that  $\tau \ll \gamma^{-1}$ in a tight-binding simulation, fewer electrons are injected into the system, resulting in $n_t \leq \bar{n}$.  In Figure \ref{fig:nt_inject} we plot the relative difference between $n_t$ and $\bar{n}$ for different values of $\tau$ and $\bar{n}$. The vertical black line marks the point where $V_p=2\gamma$, above which the bandwidth trivially affects the pulse.
Therefore, in order for the simulation to remain in the universal limit (i.e.\ unaffected by the microscopic model), the simulations must therefore remain in the
``white region'' (lower right) of Figure \ref{fig:nt_inject}. Since we have used $\tau \geq 10$ and $\bar{n} \leq 4$ throughout the article, this condition is quantitatively fulfilled except for a few points
($\tau \approx 10$ \emph{and} $\bar{n} \approx 4$) for which there are small deviations.

\subsubsection{Current conservation check}
\label{sec:appendix_current_conserve}

\begin{figure}[tbh]
	\centering
	\includegraphics[width=85mm]{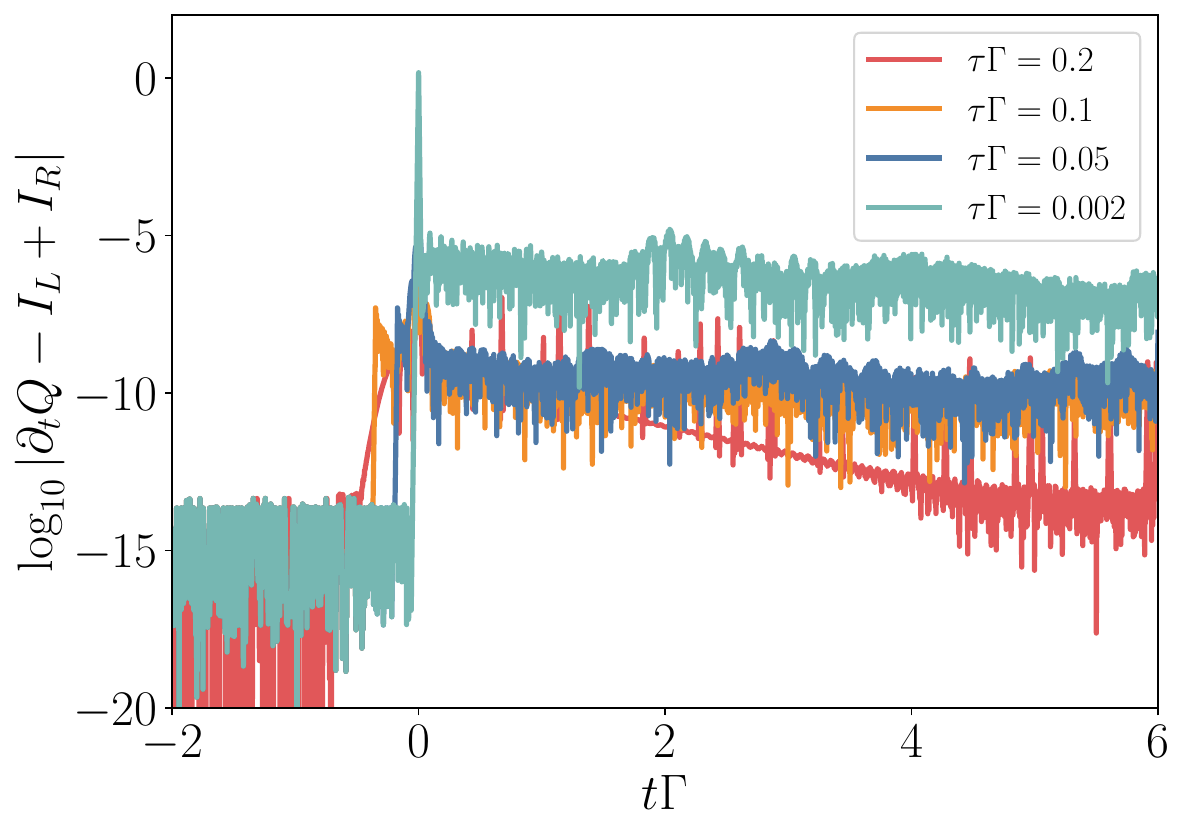}
	\vspace{-1mm}
  \caption{%
Accuracy of the current conservation Eq.\ \eqref{eq:continuity} at the impurity site for different values of the pulse width $\tau$. 
The calculation is performed for the numerical data of Fig.\ \ref{fig:dens_curr_time}, which corresponds to $\epsilon_0 = 0, \, \bar{n} = 0.75$.
}
\label{fig:current_conserve}
\end{figure}

We have checked that the current is conserved by evaluating the continuity relation Eq.\ \eqref{eq:continuity} at the impurity site.
Figure \ref{fig:current_conserve} shows the result for the numerical data from Fig.\ \ref{fig:dens_curr_time} in the main text. To evaluate $\partial_t Q$ numerically, $Q(t)$ is fitted by a cubic spline. The result in Fig.\ \ref{fig:current_conserve} shows that the current is conserved as expected, and also allows for an independent estimate of the numerical accuracy of the simulated densities and currents. It can be seen that the numerical error increases at the injection of the pulse (at time $t \Gamma = 0$), but generally remains below $10^{-5}$, except for the shortest pulse width at around $t = 0$. The numerical accuracy corresponds well to the (default) numerical accuracy used in the Tkwant software, but additionally includes the contribution from the spline interpolation of the time-discretized curves.

\subsubsection{Global check of the universal limit}
\label{sec:appendix_gamma_eff}

\begin{figure}[tbh]
	\centering
	\includegraphics[width=85mm]{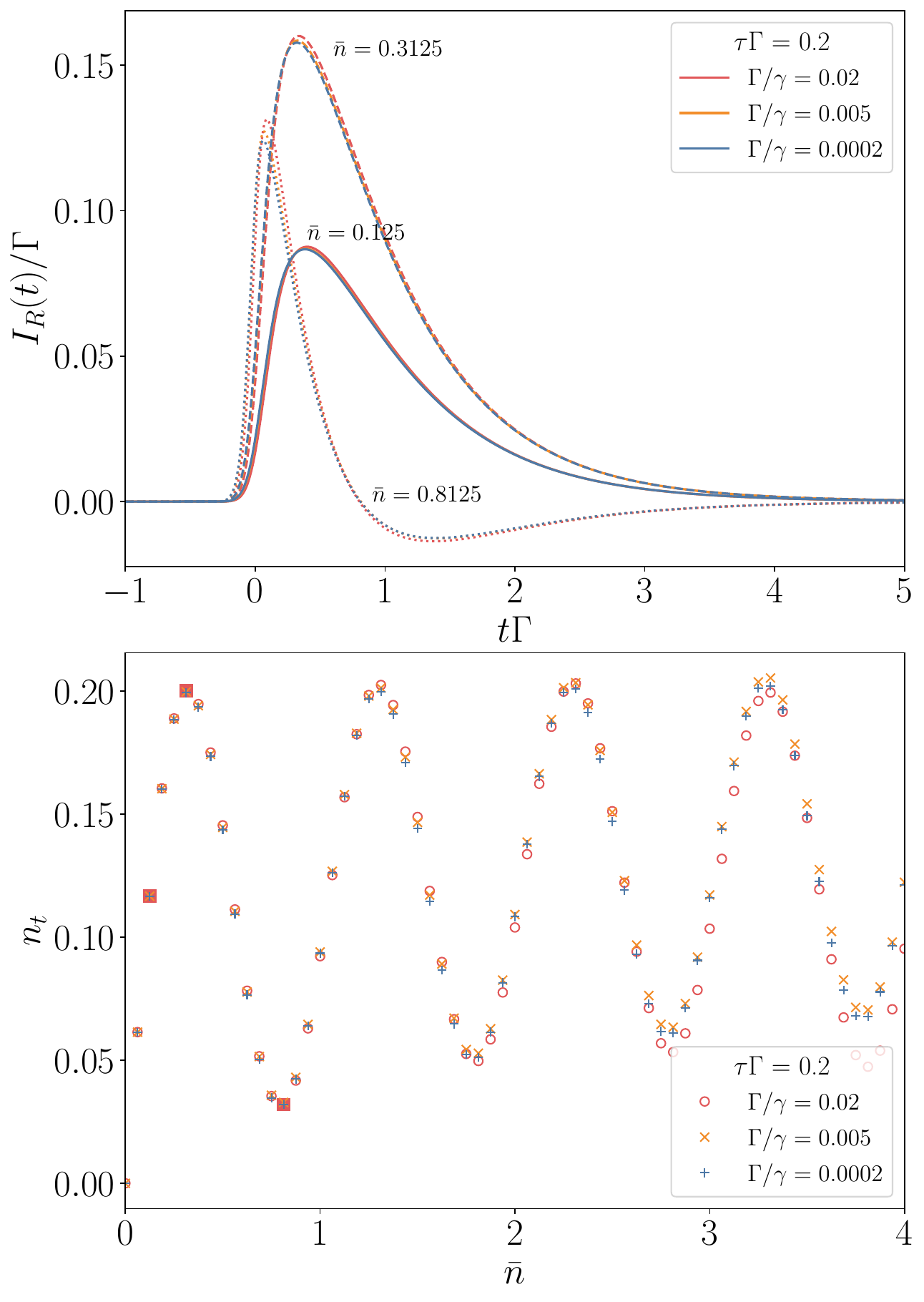}
	\vspace{-1mm}
  \caption{%
Upper panel: The properly rescaled current curves $I_R(t)$ vs.\ $t$ all collapse to a unique curve depending only on $\bar{n}$.
The data collapse is shown for three different $\bar{n}$ values: $\bar{n} = 0.125$ (solid line), $\bar{n} = 0.3125$ (dashed line) and $\bar{n} = 0.8125$ (dotted line), and the corresponding points in the lower panel are highlighted by filled red squares.  Three different value pairs of $\Gamma$ and $\tau$ are considered, such that $\tau \Gamma  = 0.2$ holds for each pair.
Lower panel: The corresponding number of transmitted charges $n_t$ as a function of $\bar{n}$ for the same $\tau \Gamma$ pairs as in the upper panel. All data are obtained from Tkwant simulations with $\epsilon_0 = 0$.
}
\label{fig:rescale_inv}
\end{figure}

We conclude this appendix with a global check that the system is in a universal (also known as wideband limit), i.e.\ that $\Gamma$ is the only energy scale describing the device.
This is done by independently varying both $\Gamma$ and $\tau$ and checking that the
results depend only on the dimensionless parameter $\tau \Gamma$.
The upper panel of Figure \ref{fig:rescale_inv} shows the actual time dependence of the transmitted current $I_R(t)$. We have run simulations for $\epsilon_0 = 0$ and for three different pairs of values of $\tau$ and $\Gamma$, such that the product is always $\tau \Gamma = 0.2$. Rescaling the time and current in units of $\Gamma$, we find that for a given value of $\tau \Gamma$, all the curves collapse into the same universal form which depends only on $\bar{n}$.
The lower panel of Figure \ref{fig:rescale_inv} shows the corresponding $n_t$ vs.\ $\bar{n}$ curves, which are obtained by integrating over the current curve according to the definition in Eq.\ \eqref{eq:nt_def}. As a guide for the eye, we have also plotted the theoretical prediction in the short-pulse limit.

\end{appendix}

\clearpage
\bibliography{meso}

\begin{thebibliography}{36}%
\makeatletter
\providecommand \@ifxundefined [1]{%
 \@ifx{#1\undefined}
}%
\providecommand \@ifnum [1]{%
 \ifnum #1\expandafter \@firstoftwo
 \else \expandafter \@secondoftwo
 \fi
}%
\providecommand \@ifx [1]{%
 \ifx #1\expandafter \@firstoftwo
 \else \expandafter \@secondoftwo
 \fi
}%
\providecommand \natexlab [1]{#1}%
\providecommand \enquote  [1]{``#1''}%
\providecommand \bibnamefont  [1]{#1}%
\providecommand \bibfnamefont [1]{#1}%
\providecommand \citenamefont [1]{#1}%
\providecommand \href@noop [0]{\@secondoftwo}%
\providecommand \href [0]{\begingroup \@sanitize@url \@href}%
\providecommand \@href[1]{\@@startlink{#1}\@@href}%
\providecommand \@@href[1]{\endgroup#1\@@endlink}%
\providecommand \@sanitize@url [0]{\catcode `\\12\catcode `\$12\catcode
  `\&12\catcode `\#12\catcode `\^12\catcode `\_12\catcode `\%12\relax}%
\providecommand \@@startlink[1]{}%
\providecommand \@@endlink[0]{}%
\providecommand \url  [0]{\begingroup\@sanitize@url \@url }%
\providecommand \@url [1]{\endgroup\@href {#1}{\urlprefix }}%
\providecommand \urlprefix  [0]{URL }%
\providecommand \Eprint [0]{\href }%
\providecommand \doibase [0]{https://doi.org/}%
\providecommand \selectlanguage [0]{\@gobble}%
\providecommand \bibinfo  [0]{\@secondoftwo}%
\providecommand \bibfield  [0]{\@secondoftwo}%
\providecommand \translation [1]{[#1]}%
\providecommand \BibitemOpen [0]{}%
\providecommand \bibitemStop [0]{}%
\providecommand \bibitemNoStop [0]{.\EOS\space}%
\providecommand \EOS [0]{\spacefactor3000\relax}%
\providecommand \BibitemShut  [1]{\csname bibitem#1\endcsname}%
\let\auto@bib@innerbib\@empty
\bibitem [{\citenamefont {Levitov}\ \emph {et~al.}(1996)\citenamefont
  {Levitov}, \citenamefont {Lee},\ and\ \citenamefont {Lesovik}}]{levitov96}%
  \BibitemOpen
  \bibfield  {author} {\bibinfo {author} {\bibfnamefont {L.~S.}\ \bibnamefont
  {Levitov}}, \bibinfo {author} {\bibfnamefont {H.}~\bibnamefont {Lee}},\ and\
  \bibinfo {author} {\bibfnamefont {G.~B.}\ \bibnamefont {Lesovik}},\
  }\bibfield  {title} {\bibinfo {title} {Electron counting statistics and
  coherent states of electric current},\ }\href
  {https://doi.org/10.1063/1.531672} {\bibfield  {journal} {\bibinfo  {journal}
  {J. Math. Phys.}\ }\textbf {\bibinfo {volume} {37}},\ \bibinfo {pages} {4845}
  (\bibinfo {year} {1996})}\BibitemShut {NoStop}%
\bibitem [{\citenamefont {Ivanov}\ \emph {et~al.}(1997)\citenamefont {Ivanov},
  \citenamefont {Lee},\ and\ \citenamefont {Levitov}}]{levitov97}%
  \BibitemOpen
  \bibfield  {author} {\bibinfo {author} {\bibfnamefont {D.~A.}\ \bibnamefont
  {Ivanov}}, \bibinfo {author} {\bibfnamefont {H.~W.}\ \bibnamefont {Lee}},\
  and\ \bibinfo {author} {\bibfnamefont {L.~S.}\ \bibnamefont {Levitov}},\
  }\bibfield  {title} {\bibinfo {title} {Coherent states of alternating
  current},\ }\href {https://doi.org/10.1103/PhysRevB.56.6839} {\bibfield
  {journal} {\bibinfo  {journal} {Phys. Rev. B}\ }\textbf {\bibinfo {volume}
  {56}},\ \bibinfo {pages} {6839} (\bibinfo {year} {1997})}\BibitemShut
  {NoStop}%
\bibitem [{\citenamefont {Keeling}\ \emph {et~al.}(2006)\citenamefont
  {Keeling}, \citenamefont {Klich},\ and\ \citenamefont {Levitov}}]{keeling06}%
  \BibitemOpen
  \bibfield  {author} {\bibinfo {author} {\bibfnamefont {J.}~\bibnamefont
  {Keeling}}, \bibinfo {author} {\bibfnamefont {I.}~\bibnamefont {Klich}},\
  and\ \bibinfo {author} {\bibfnamefont {L.~S.}\ \bibnamefont {Levitov}},\
  }\bibfield  {title} {\bibinfo {title} {Minimal excitation states of electrons
  in one-dimensional wires},\ }\href
  {https://doi.org/10.1103/PhysRevLett.97.116403} {\bibfield  {journal}
  {\bibinfo  {journal} {Phys. Rev. Lett.}\ }\textbf {\bibinfo {volume} {97}},\
  \bibinfo {pages} {116403} (\bibinfo {year} {2006})}\BibitemShut {NoStop}%
\bibitem [{\citenamefont {Dubois}\ \emph {et~al.}(2013)\citenamefont {Dubois},
  \citenamefont {Jullien}, \citenamefont {Portier}, \citenamefont {Roche},
  \citenamefont {Cavanna}, \citenamefont {Jin}, \citenamefont {Wegscheider},
  \citenamefont {Roulleau},\ and\ \citenamefont {Glattli}}]{Dubois13}%
  \BibitemOpen
  \bibfield  {author} {\bibinfo {author} {\bibfnamefont {J.}~\bibnamefont
  {Dubois}}, \bibinfo {author} {\bibfnamefont {T.}~\bibnamefont {Jullien}},
  \bibinfo {author} {\bibfnamefont {F.}~\bibnamefont {Portier}}, \bibinfo
  {author} {\bibfnamefont {P.}~\bibnamefont {Roche}}, \bibinfo {author}
  {\bibfnamefont {A.}~\bibnamefont {Cavanna}}, \bibinfo {author} {\bibfnamefont
  {Y.}~\bibnamefont {Jin}}, \bibinfo {author} {\bibfnamefont {W.}~\bibnamefont
  {Wegscheider}}, \bibinfo {author} {\bibfnamefont {P.}~\bibnamefont
  {Roulleau}},\ and\ \bibinfo {author} {\bibfnamefont {D.}~\bibnamefont
  {Glattli}},\ }\bibfield  {title} {\bibinfo {title} {Minimal-excitation states
  for electron quantum optics using levitons},\ }\href
  {https://doi.org/10.1038/nature12713} {\bibfield  {journal} {\bibinfo
  {journal} {Nature}\ }\textbf {\bibinfo {volume} {502}},\ \bibinfo {pages}
  {659} (\bibinfo {year} {2013})}\BibitemShut {NoStop}%
\bibitem [{\citenamefont {Jullien}\ \emph {et~al.}(2014)\citenamefont
  {Jullien}, \citenamefont {Roulleau}, \citenamefont {Roche}, \citenamefont
  {Cavanna}, \citenamefont {Jin},\ and\ \citenamefont {Glattli}}]{Jullien14}%
  \BibitemOpen
  \bibfield  {author} {\bibinfo {author} {\bibfnamefont {T.}~\bibnamefont
  {Jullien}}, \bibinfo {author} {\bibfnamefont {P.}~\bibnamefont {Roulleau}},
  \bibinfo {author} {\bibfnamefont {B.}~\bibnamefont {Roche}}, \bibinfo
  {author} {\bibfnamefont {A.}~\bibnamefont {Cavanna}}, \bibinfo {author}
  {\bibfnamefont {Y.}~\bibnamefont {Jin}},\ and\ \bibinfo {author}
  {\bibfnamefont {C.~D.}\ \bibnamefont {Glattli}},\ }\bibfield  {title}
  {\bibinfo {title} {{Quantum tomography of an electron}},\ }\href
  {https://doi.org/10.1038/nature13821} {\bibfield  {journal} {\bibinfo
  {journal} {{Nature}}\ }\textbf {\bibinfo {volume} {514}},\ \bibinfo {pages}
  {603 } (\bibinfo {year} {2014})}\BibitemShut {NoStop}%
\bibitem [{\citenamefont {B{\"a}uerle}\ \emph {et~al.}(2018)\citenamefont
  {B{\"a}uerle}, \citenamefont {Glattli}, \citenamefont {Meunier},
  \citenamefont {Portier}, \citenamefont {Roche}, \citenamefont {Roulleau},
  \citenamefont {Takada},\ and\ \citenamefont {Waintal}}]{Bauerle18}%
  \BibitemOpen
  \bibfield  {author} {\bibinfo {author} {\bibfnamefont {C.}~\bibnamefont
  {B{\"a}uerle}}, \bibinfo {author} {\bibfnamefont {D.~C.}\ \bibnamefont
  {Glattli}}, \bibinfo {author} {\bibfnamefont {T.}~\bibnamefont {Meunier}},
  \bibinfo {author} {\bibfnamefont {F.}~\bibnamefont {Portier}}, \bibinfo
  {author} {\bibfnamefont {P.}~\bibnamefont {Roche}}, \bibinfo {author}
  {\bibfnamefont {P.}~\bibnamefont {Roulleau}}, \bibinfo {author}
  {\bibfnamefont {S.}~\bibnamefont {Takada}},\ and\ \bibinfo {author}
  {\bibfnamefont {X.}~\bibnamefont {Waintal}},\ }\bibfield  {title} {\bibinfo
  {title} {Coherent control of single electrons: a review of current
  progress},\ }\href {https://doi.org/10.1088/1361-6633/aaa98a} {\bibfield
  {journal} {\bibinfo  {journal} {Rep. Prog. Phys.}\ }\textbf {\bibinfo
  {volume} {81}},\ \bibinfo {pages} {056503} (\bibinfo {year}
  {2018})}\BibitemShut {NoStop}%
\bibitem [{\citenamefont {Roussel}\ \emph {et~al.}(2021)\citenamefont
  {Roussel}, \citenamefont {Cabart}, \citenamefont {F\`eve},\ and\
  \citenamefont {Degiovanni}}]{Roussel21}%
  \BibitemOpen
  \bibfield  {author} {\bibinfo {author} {\bibfnamefont {B.}~\bibnamefont
  {Roussel}}, \bibinfo {author} {\bibfnamefont {C.}~\bibnamefont {Cabart}},
  \bibinfo {author} {\bibfnamefont {G.}~\bibnamefont {F\`eve}},\ and\ \bibinfo
  {author} {\bibfnamefont {P.}~\bibnamefont {Degiovanni}},\ }\bibfield  {title}
  {\bibinfo {title} {{Processing Quantum Signals Carried by Electrical
  Currents}},\ }\href {https://doi.org/10.1103/PRXQuantum.2.020314} {\bibfield
  {journal} {\bibinfo  {journal} {PRX Quantum}\ }\textbf {\bibinfo {volume}
  {2}},\ \bibinfo {pages} {020314} (\bibinfo {year} {2021})}\BibitemShut
  {NoStop}%
\bibitem [{\citenamefont {Bartolomei}\ \emph {et~al.}(2025)\citenamefont
  {Bartolomei}, \citenamefont {Frigerio}, \citenamefont {Ruelle}, \citenamefont
  {Rebora}, \citenamefont {Jin}, \citenamefont {Gennser}, \citenamefont
  {Cavanna}, \citenamefont {Baudin}, \citenamefont {Berroir}, \citenamefont
  {Safi}, \citenamefont {Degiovanni}, \citenamefont {M\'enard},\ and\
  \citenamefont {F\`eve}}]{bartolomei25}%
  \BibitemOpen
  \bibfield  {author} {\bibinfo {author} {\bibfnamefont {H.}~\bibnamefont
  {Bartolomei}}, \bibinfo {author} {\bibfnamefont {E.}~\bibnamefont
  {Frigerio}}, \bibinfo {author} {\bibfnamefont {M.}~\bibnamefont {Ruelle}},
  \bibinfo {author} {\bibfnamefont {G.}~\bibnamefont {Rebora}}, \bibinfo
  {author} {\bibfnamefont {Y.}~\bibnamefont {Jin}}, \bibinfo {author}
  {\bibfnamefont {U.}~\bibnamefont {Gennser}}, \bibinfo {author} {\bibfnamefont
  {A.}~\bibnamefont {Cavanna}}, \bibinfo {author} {\bibfnamefont
  {E.}~\bibnamefont {Baudin}}, \bibinfo {author} {\bibfnamefont {J.-M.}\
  \bibnamefont {Berroir}}, \bibinfo {author} {\bibfnamefont {I.}~\bibnamefont
  {Safi}}, \bibinfo {author} {\bibfnamefont {P.}~\bibnamefont {Degiovanni}},
  \bibinfo {author} {\bibfnamefont {G.~C.}\ \bibnamefont {M\'enard}},\ and\
  \bibinfo {author} {\bibfnamefont {G.}~\bibnamefont {F\`eve}},\ }\bibfield
  {title} {\bibinfo {title} {Time-resolved sensing of electromagnetic fields
  with single-electron interferometry},\ }\bibfield  {journal} {\bibinfo
  {journal} {Nat. Nanotechnol.}\ }\href
  {https://doi.org/10.1038/s41565-025-01888-2} {10.1038/s41565-025-01888-2}
  (\bibinfo {year} {2025})\BibitemShut {NoStop}%
\bibitem [{\citenamefont {Souquet-Basiège}\ \emph {et~al.}()\citenamefont
  {Souquet-Basiège}, \citenamefont {Roussel}, \citenamefont {Rebora},
  \citenamefont {M\'enard}, \citenamefont {Safi}, \citenamefont {F\`eve},\ and\
  \citenamefont {Degiovanni}}]{souquetbasiege24}%
  \BibitemOpen
  \bibfield  {author} {\bibinfo {author} {\bibfnamefont {H.}~\bibnamefont
  {Souquet-Basiège}}, \bibinfo {author} {\bibfnamefont {B.}~\bibnamefont
  {Roussel}}, \bibinfo {author} {\bibfnamefont {G.}~\bibnamefont {Rebora}},
  \bibinfo {author} {\bibfnamefont {G.}~\bibnamefont {M\'enard}}, \bibinfo
  {author} {\bibfnamefont {I.}~\bibnamefont {Safi}}, \bibinfo {author}
  {\bibfnamefont {G.}~\bibnamefont {F\`eve}},\ and\ \bibinfo {author}
  {\bibfnamefont {P.}~\bibnamefont {Degiovanni}},\ }\href
  {https://arxiv.org/abs/2405.05796} {\bibinfo {title} {Quantum sensing of time
  dependent electromagnetic fields with single electron excitations}},\ \Eprint
  {https://arxiv.org/abs/2405.05796} {arXiv:2405.05796 [quant-ph]} \BibitemShut
  {NoStop}%
\bibitem [{\citenamefont {Ouacel}\ \emph {et~al.}(2025)\citenamefont {Ouacel},
  \citenamefont {Mazzella}, \citenamefont {Kloss}, \citenamefont {Aluffi},
  \citenamefont {Vasselon}, \citenamefont {Edlbauer}, \citenamefont {Wang},
  \citenamefont {Geffroy}, \citenamefont {Shaju}, \citenamefont {Ludwig},
  \citenamefont {Wieck}, \citenamefont {Yamamoto}, \citenamefont {Pomaranski},
  \citenamefont {Takada}, \citenamefont {Kaneko}, \citenamefont {Georgiou},
  \citenamefont {Waintal}, \citenamefont {Urdampilleta}, \citenamefont
  {Sellier},\ and\ \citenamefont {B{\"a}uerle}}]{ouacel2025}%
  \BibitemOpen
  \bibfield  {author} {\bibinfo {author} {\bibfnamefont {S.}~\bibnamefont
  {Ouacel}}, \bibinfo {author} {\bibfnamefont {L.}~\bibnamefont {Mazzella}},
  \bibinfo {author} {\bibfnamefont {T.}~\bibnamefont {Kloss}}, \bibinfo
  {author} {\bibfnamefont {M.}~\bibnamefont {Aluffi}}, \bibinfo {author}
  {\bibfnamefont {T.}~\bibnamefont {Vasselon}}, \bibinfo {author}
  {\bibfnamefont {H.}~\bibnamefont {Edlbauer}}, \bibinfo {author}
  {\bibfnamefont {J.}~\bibnamefont {Wang}}, \bibinfo {author} {\bibfnamefont
  {C.}~\bibnamefont {Geffroy}}, \bibinfo {author} {\bibfnamefont
  {J.}~\bibnamefont {Shaju}}, \bibinfo {author} {\bibfnamefont
  {A.}~\bibnamefont {Ludwig}}, \bibinfo {author} {\bibfnamefont {A.~D.}\
  \bibnamefont {Wieck}}, \bibinfo {author} {\bibfnamefont {M.}~\bibnamefont
  {Yamamoto}}, \bibinfo {author} {\bibfnamefont {D.}~\bibnamefont
  {Pomaranski}}, \bibinfo {author} {\bibfnamefont {S.}~\bibnamefont {Takada}},
  \bibinfo {author} {\bibfnamefont {N.-H.}\ \bibnamefont {Kaneko}}, \bibinfo
  {author} {\bibfnamefont {G.}~\bibnamefont {Georgiou}}, \bibinfo {author}
  {\bibfnamefont {X.}~\bibnamefont {Waintal}}, \bibinfo {author} {\bibfnamefont
  {M.}~\bibnamefont {Urdampilleta}}, \bibinfo {author} {\bibfnamefont
  {H.}~\bibnamefont {Sellier}},\ and\ \bibinfo {author} {\bibfnamefont
  {C.}~\bibnamefont {B{\"a}uerle}},\ }\bibfield  {title} {\bibinfo {title}
  {Electronic interferometry with ultrashort plasmonic pulses},\ }\href
  {https://doi.org/10.1038/s41467-025-58939-4} {\bibfield  {journal} {\bibinfo
  {journal} {{Nat. Commun.}}\ }\textbf {\bibinfo {volume} {16}},\ \bibinfo
  {pages} {4632} (\bibinfo {year} {2025})}\BibitemShut {NoStop}%
\bibitem [{\citenamefont {Assouline}\ \emph {et~al.}(2023)\citenamefont
  {Assouline}, \citenamefont {Pugliese}, \citenamefont {Chakraborti},
  \citenamefont {Lee}, \citenamefont {Bernabeu}, \citenamefont {Jo},
  \citenamefont {Watanabe}, \citenamefont {Taniguchi}, \citenamefont {Glattli},
  \citenamefont {Kumada}, \citenamefont {Sim}, \citenamefont {Parmentier},\
  and\ \citenamefont {Roulleau}}]{Assouline23}%
  \BibitemOpen
  \bibfield  {author} {\bibinfo {author} {\bibfnamefont {A.}~\bibnamefont
  {Assouline}}, \bibinfo {author} {\bibfnamefont {L.}~\bibnamefont {Pugliese}},
  \bibinfo {author} {\bibfnamefont {H.}~\bibnamefont {Chakraborti}}, \bibinfo
  {author} {\bibfnamefont {S.}~\bibnamefont {Lee}}, \bibinfo {author}
  {\bibfnamefont {L.}~\bibnamefont {Bernabeu}}, \bibinfo {author}
  {\bibfnamefont {M.}~\bibnamefont {Jo}}, \bibinfo {author} {\bibfnamefont
  {K.}~\bibnamefont {Watanabe}}, \bibinfo {author} {\bibfnamefont
  {T.}~\bibnamefont {Taniguchi}}, \bibinfo {author} {\bibfnamefont {D.~C.}\
  \bibnamefont {Glattli}}, \bibinfo {author} {\bibfnamefont {N.}~\bibnamefont
  {Kumada}}, \bibinfo {author} {\bibfnamefont {H.-S.}\ \bibnamefont {Sim}},
  \bibinfo {author} {\bibfnamefont {F.~D.}\ \bibnamefont {Parmentier}},\ and\
  \bibinfo {author} {\bibfnamefont {P.}~\bibnamefont {Roulleau}},\ }\bibfield
  {title} {\bibinfo {title} {{Emission and coherent control of Levitons in
  graphene}},\ }\href {https://doi.org/10.1126/science.adf9887} {\bibfield
  {journal} {\bibinfo  {journal} {Science}\ }\textbf {\bibinfo {volume}
  {382}},\ \bibinfo {pages} {1260} (\bibinfo {year} {2023})}\BibitemShut
  {NoStop}%
\bibitem [{\citenamefont {Gaury}\ and\ \citenamefont
  {Waintal}(2014)}]{gaury14a}%
  \BibitemOpen
  \bibfield  {author} {\bibinfo {author} {\bibfnamefont {B.}~\bibnamefont
  {Gaury}}\ and\ \bibinfo {author} {\bibfnamefont {X.}~\bibnamefont
  {Waintal}},\ }\bibfield  {title} {\bibinfo {title} {Dynamical control of
  interference using voltage pulses in the quantum regime},\ }\href
  {https://doi.org/10.1038/ncomms4844} {\bibfield  {journal} {\bibinfo
  {journal} {Nat. Commun.}\ }\textbf {\bibinfo {volume} {5}},\ \bibinfo {pages}
  {3844} (\bibinfo {year} {2014})}\BibitemShut {NoStop}%
\bibitem [{\citenamefont {Gaury}\ \emph
  {et~al.}(2014{\natexlab{a}})\citenamefont {Gaury}, \citenamefont {Weston},\
  and\ \citenamefont {Waintal}}]{gaury14b}%
  \BibitemOpen
  \bibfield  {author} {\bibinfo {author} {\bibfnamefont {B.}~\bibnamefont
  {Gaury}}, \bibinfo {author} {\bibfnamefont {J.}~\bibnamefont {Weston}},\ and\
  \bibinfo {author} {\bibfnamefont {X.}~\bibnamefont {Waintal}},\ }\bibfield
  {title} {\bibinfo {title} {{Stopping electrons with radio-frequency pulses in
  the quantum Hall regime}},\ }\href
  {https://doi.org/10.1103/PhysRevB.90.161305} {\bibfield  {journal} {\bibinfo
  {journal} {Phys. Rev. B}\ }\textbf {\bibinfo {volume} {90}},\ \bibinfo
  {pages} {161305} (\bibinfo {year} {2014}{\natexlab{a}})}\BibitemShut
  {NoStop}%
\bibitem [{\citenamefont {Gaury}\ \emph {et~al.}(2015)\citenamefont {Gaury},
  \citenamefont {Weston},\ and\ \citenamefont {Waintal}}]{gaury15}%
  \BibitemOpen
  \bibfield  {author} {\bibinfo {author} {\bibfnamefont {B.}~\bibnamefont
  {Gaury}}, \bibinfo {author} {\bibfnamefont {J.}~\bibnamefont {Weston}},\ and\
  \bibinfo {author} {\bibfnamefont {X.}~\bibnamefont {Waintal}},\ }\bibfield
  {title} {\bibinfo {title} {{The a.c. Josephson effect without
  superconductivity}},\ }\href {https://doi.org/10.1038/ncomms7524} {\bibfield
  {journal} {\bibinfo  {journal} {Nat. Commun.}\ }\textbf {\bibinfo {volume}
  {6}},\ \bibinfo {pages} {6524} (\bibinfo {year} {2015})}\BibitemShut
  {NoStop}%
\bibitem [{\citenamefont {Weston}\ \emph {et~al.}(2015)\citenamefont {Weston},
  \citenamefont {Gaury},\ and\ \citenamefont {Waintal}}]{weston15}%
  \BibitemOpen
  \bibfield  {author} {\bibinfo {author} {\bibfnamefont {J.}~\bibnamefont
  {Weston}}, \bibinfo {author} {\bibfnamefont {B.}~\bibnamefont {Gaury}},\ and\
  \bibinfo {author} {\bibfnamefont {X.}~\bibnamefont {Waintal}},\ }\bibfield
  {title} {\bibinfo {title} {{Manipulating Andreev and Majorana bound states
  with microwaves}},\ }\href {https://doi.org/10.1103/PhysRevB.92.020513}
  {\bibfield  {journal} {\bibinfo  {journal} {Phys. Rev. B}\ }\textbf {\bibinfo
  {volume} {92}},\ \bibinfo {pages} {020513} (\bibinfo {year}
  {2015})}\BibitemShut {NoStop}%
\bibitem [{\citenamefont {Meyer}\ \emph {et~al.}(2017)\citenamefont {Meyer},
  \citenamefont {Haack}, \citenamefont {Groth},\ and\ \citenamefont
  {Waintal}}]{meyer17}%
  \BibitemOpen
  \bibfield  {author} {\bibinfo {author} {\bibfnamefont {U.}~\bibnamefont
  {Meyer}}, \bibinfo {author} {\bibfnamefont {G.}~\bibnamefont {Haack}},
  \bibinfo {author} {\bibfnamefont {C.}~\bibnamefont {Groth}},\ and\ \bibinfo
  {author} {\bibfnamefont {X.}~\bibnamefont {Waintal}},\ }\bibfield  {title}
  {\bibinfo {title} {{Control of the Oscillatory Interlayer Exchange
  Interaction with Terahertz Radiation}},\ }\href
  {https://doi.org/10.1103/PhysRevLett.118.097701} {\bibfield  {journal}
  {\bibinfo  {journal} {Phys. Rev. Lett.}\ }\textbf {\bibinfo {volume} {118}},\
  \bibinfo {pages} {097701} (\bibinfo {year} {2017})}\BibitemShut {NoStop}%
\bibitem [{\citenamefont {Kouwenhoven}\ \emph {et~al.}(1997)\citenamefont
  {Kouwenhoven}, \citenamefont {Marcus}, \citenamefont {McEuen}, \citenamefont
  {Tarucha}, \citenamefont {Westervelt},\ and\ \citenamefont
  {Wingreen}}]{Kouwenhoven97}%
  \BibitemOpen
  \bibfield  {author} {\bibinfo {author} {\bibfnamefont {L.~P.}\ \bibnamefont
  {Kouwenhoven}}, \bibinfo {author} {\bibfnamefont {C.~M.}\ \bibnamefont
  {Marcus}}, \bibinfo {author} {\bibfnamefont {P.~L.}\ \bibnamefont {McEuen}},
  \bibinfo {author} {\bibfnamefont {S.}~\bibnamefont {Tarucha}}, \bibinfo
  {author} {\bibfnamefont {R.~M.}\ \bibnamefont {Westervelt}},\ and\ \bibinfo
  {author} {\bibfnamefont {N.~S.}\ \bibnamefont {Wingreen}},\ }\bibinfo {title}
  {Electron transport in quantum dots},\ in\ \href
  {https://doi.org/10.1007/978-94-015-8839-3_4} {\emph {\bibinfo {booktitle}
  {Mesoscopic Electron Transport}}},\ \bibinfo {editor} {edited by\ \bibinfo
  {editor} {\bibfnamefont {L.~L.}\ \bibnamefont {Sohn}}, \bibinfo {editor}
  {\bibfnamefont {L.~P.}\ \bibnamefont {Kouwenhoven}},\ and\ \bibinfo {editor}
  {\bibfnamefont {G.}~\bibnamefont {Sch{\"o}n}}}\ (\bibinfo  {publisher}
  {Springer Netherlands},\ \bibinfo {address} {Dordrecht},\ \bibinfo {year}
  {1997})\ pp.\ \bibinfo {pages} {105--214}\BibitemShut {NoStop}%
\bibitem [{\citenamefont {Gaury}\ \emph
  {et~al.}(2014{\natexlab{b}})\citenamefont {Gaury}, \citenamefont {Weston},
  \citenamefont {Santin}, \citenamefont {Houzet}, \citenamefont {Groth},\ and\
  \citenamefont {Waintal}}]{gaury14}%
  \BibitemOpen
  \bibfield  {author} {\bibinfo {author} {\bibfnamefont {B.}~\bibnamefont
  {Gaury}}, \bibinfo {author} {\bibfnamefont {J.}~\bibnamefont {Weston}},
  \bibinfo {author} {\bibfnamefont {M.}~\bibnamefont {Santin}}, \bibinfo
  {author} {\bibfnamefont {M.}~\bibnamefont {Houzet}}, \bibinfo {author}
  {\bibfnamefont {C.}~\bibnamefont {Groth}},\ and\ \bibinfo {author}
  {\bibfnamefont {X.}~\bibnamefont {Waintal}},\ }\bibfield  {title} {\bibinfo
  {title} {Numerical simulations of time-resolved quantum electronics},\ }\href
  {https://doi.org/http://dx.doi.org/10.1016/j.physrep.2013.09.001} {\bibfield
  {journal} {\bibinfo  {journal} {Phys. Rep.}\ }\textbf {\bibinfo {volume}
  {534}},\ \bibinfo {pages} {1 } (\bibinfo {year}
  {2014}{\natexlab{b}})}\BibitemShut {NoStop}%
\bibitem [{\citenamefont {Kloss}\ \emph {et~al.}(2021)\citenamefont {Kloss},
  \citenamefont {Weston}, \citenamefont {Gaury}, \citenamefont {Rossignol},
  \citenamefont {Groth},\ and\ \citenamefont {Waintal}}]{Tkwant21}%
  \BibitemOpen
  \bibfield  {author} {\bibinfo {author} {\bibfnamefont {T.}~\bibnamefont
  {Kloss}}, \bibinfo {author} {\bibfnamefont {J.}~\bibnamefont {Weston}},
  \bibinfo {author} {\bibfnamefont {B.}~\bibnamefont {Gaury}}, \bibinfo
  {author} {\bibfnamefont {B.}~\bibnamefont {Rossignol}}, \bibinfo {author}
  {\bibfnamefont {C.}~\bibnamefont {Groth}},\ and\ \bibinfo {author}
  {\bibfnamefont {X.}~\bibnamefont {Waintal}},\ }\bibfield  {title} {\bibinfo
  {title} {Tkwant: a software package for time-dependent quantum transport},\
  }\href {https://doi.org/10.1088/1367-2630/abddf7} {\bibfield  {journal}
  {\bibinfo  {journal} {New J. Phys.}\ }\textbf {\bibinfo {volume} {23}},\
  \bibinfo {pages} {023025} (\bibinfo {year} {2021})}\BibitemShut {NoStop}%
\bibitem [{\citenamefont {Abramowitz}\ and\ \citenamefont
  {Stegun}(1972)}]{AbraSteg72}%
  \BibitemOpen
  \bibinfo {editor} {\bibfnamefont {M.}~\bibnamefont {Abramowitz}}\ and\
  \bibinfo {editor} {\bibfnamefont {I.~A.}\ \bibnamefont {Stegun}},\ eds.,\
  \href@noop {} {\emph {\bibinfo {title} {Handbook of Mathematical Functions
  with Formulas, Graphs, and Mathematical Tables}}}\ (\bibinfo  {publisher}
  {New York: Dover Publications},\ \bibinfo {year} {1972})\BibitemShut
  {NoStop}%
\bibitem [{tkw()}]{tkwant}%
  \BibitemOpen
  \href@noop {} {}\bibinfo {note} {\textsc{Tkwant} is free software and can be
  found at
  \hyperlink{https://tkwant.kwant-project.org}{https://tkwant.kwant-project.org}}\BibitemShut
  {NoStop}%
\bibitem [{\citenamefont {Buttiker}(1993)}]{Buttiker93}%
  \BibitemOpen
  \bibfield  {author} {\bibinfo {author} {\bibfnamefont {M.}~\bibnamefont
  {Buttiker}},\ }\bibfield  {title} {\bibinfo {title} {Capacitance, admittance,
  and rectification properties of small conductors},\ }\href
  {https://doi.org/10.1088/0953-8984/5/50/017} {\bibfield  {journal} {\bibinfo
  {journal} {J. Phys.: Condens. Matter}\ }\textbf {\bibinfo {volume} {5}},\
  \bibinfo {pages} {9361} (\bibinfo {year} {1993})}\BibitemShut {NoStop}%
\bibitem [{\citenamefont {Nakamura}\ \emph {et~al.}(2020)\citenamefont
  {Nakamura}, \citenamefont {Liang}, \citenamefont {Gardner},\ and\
  \citenamefont {Manfra}}]{Manfra20}%
  \BibitemOpen
  \bibfield  {author} {\bibinfo {author} {\bibfnamefont {J.}~\bibnamefont
  {Nakamura}}, \bibinfo {author} {\bibfnamefont {S.}~\bibnamefont {Liang}},
  \bibinfo {author} {\bibfnamefont {G.~C.}\ \bibnamefont {Gardner}},\ and\
  \bibinfo {author} {\bibfnamefont {M.~J.}\ \bibnamefont {Manfra}},\ }\bibfield
   {title} {\bibinfo {title} {Direct observation of anyonic braiding
  statistics},\ }\href {https://doi.org/10.1038/s41567-020-1019-1} {\bibfield
  {journal} {\bibinfo  {journal} {Nature Physics}\ }\textbf {\bibinfo {volume}
  {16}},\ \bibinfo {pages} {931} (\bibinfo {year} {2020})}\BibitemShut
  {NoStop}%
\bibitem [{\citenamefont {Ronen}\ \emph {et~al.}(2021)\citenamefont {Ronen},
  \citenamefont {Werkmeister}, \citenamefont {Haie~Najafabadi}, \citenamefont
  {Pierce}, \citenamefont {Anderson}, \citenamefont {Shin}, \citenamefont
  {Lee}, \citenamefont {Lee}, \citenamefont {Johnson}, \citenamefont
  {Watanabe}, \citenamefont {Taniguchi}, \citenamefont {Yacoby},\ and\
  \citenamefont {Kim}}]{ronen21}%
  \BibitemOpen
  \bibfield  {author} {\bibinfo {author} {\bibfnamefont {Y.}~\bibnamefont
  {Ronen}}, \bibinfo {author} {\bibfnamefont {T.}~\bibnamefont {Werkmeister}},
  \bibinfo {author} {\bibfnamefont {D.}~\bibnamefont {Haie~Najafabadi}},
  \bibinfo {author} {\bibfnamefont {A.~T.}\ \bibnamefont {Pierce}}, \bibinfo
  {author} {\bibfnamefont {L.~E.}\ \bibnamefont {Anderson}}, \bibinfo {author}
  {\bibfnamefont {Y.~J.}\ \bibnamefont {Shin}}, \bibinfo {author}
  {\bibfnamefont {S.~Y.}\ \bibnamefont {Lee}}, \bibinfo {author} {\bibfnamefont
  {Y.~H.}\ \bibnamefont {Lee}}, \bibinfo {author} {\bibfnamefont
  {B.}~\bibnamefont {Johnson}}, \bibinfo {author} {\bibfnamefont
  {K.}~\bibnamefont {Watanabe}}, \bibinfo {author} {\bibfnamefont
  {T.}~\bibnamefont {Taniguchi}}, \bibinfo {author} {\bibfnamefont
  {A.}~\bibnamefont {Yacoby}},\ and\ \bibinfo {author} {\bibfnamefont
  {P.}~\bibnamefont {Kim}},\ }\bibfield  {title} {\bibinfo {title}
  {Aharonov–{Bohm} effect in graphene-based {Fabry}–{Pérot} quantum {Hall}
  interferometers},\ }\href {https://doi.org/10.1038/s41565-021-00861-z}
  {\bibfield  {journal} {\bibinfo  {journal} {Nat. Nanotechnol.}\ }\textbf
  {\bibinfo {volume} {16}},\ \bibinfo {pages} {563} (\bibinfo {year}
  {2021})}\BibitemShut {NoStop}%
\bibitem [{\citenamefont {Fujisawa}\ \emph {et~al.}(2006)\citenamefont
  {Fujisawa}, \citenamefont {Hayashi},\ and\ \citenamefont
  {Sasaki}}]{Fujisawa06}%
  \BibitemOpen
  \bibfield  {author} {\bibinfo {author} {\bibfnamefont {T.}~\bibnamefont
  {Fujisawa}}, \bibinfo {author} {\bibfnamefont {T.}~\bibnamefont {Hayashi}},\
  and\ \bibinfo {author} {\bibfnamefont {S.}~\bibnamefont {Sasaki}},\
  }\bibfield  {title} {\bibinfo {title} {Time-dependent single-electron
  transport through quantum dots},\ }\href
  {https://doi.org/10.1088/0034-4885/69/3/R05} {\bibfield  {journal} {\bibinfo
  {journal} {Rep. Prog. Phys.}\ }\textbf {\bibinfo {volume} {69}},\ \bibinfo
  {pages} {759} (\bibinfo {year} {2006})}\BibitemShut {NoStop}%
\bibitem [{\citenamefont {Uimonen}\ \emph {et~al.}(2011)\citenamefont
  {Uimonen}, \citenamefont {Khosravi}, \citenamefont {Stan}, \citenamefont
  {Stefanucci}, \citenamefont {Kurth}, \citenamefont {van Leeuwen},\ and\
  \citenamefont {Gross}}]{Uimonen11}%
  \BibitemOpen
  \bibfield  {author} {\bibinfo {author} {\bibfnamefont {A.-M.}\ \bibnamefont
  {Uimonen}}, \bibinfo {author} {\bibfnamefont {E.}~\bibnamefont {Khosravi}},
  \bibinfo {author} {\bibfnamefont {A.}~\bibnamefont {Stan}}, \bibinfo {author}
  {\bibfnamefont {G.}~\bibnamefont {Stefanucci}}, \bibinfo {author}
  {\bibfnamefont {S.}~\bibnamefont {Kurth}}, \bibinfo {author} {\bibfnamefont
  {R.}~\bibnamefont {van Leeuwen}},\ and\ \bibinfo {author} {\bibfnamefont
  {E.~K.~U.}\ \bibnamefont {Gross}},\ }\bibfield  {title} {\bibinfo {title}
  {{Comparative study of many-body perturbation theory and time-dependent
  density functional theory in the out-of-equilibrium Anderson model}},\ }\href
  {https://doi.org/10.1103/PhysRevB.84.115103} {\bibfield  {journal} {\bibinfo
  {journal} {Phys. Rev. B}\ }\textbf {\bibinfo {volume} {84}},\ \bibinfo
  {pages} {115103} (\bibinfo {year} {2011})}\BibitemShut {NoStop}%
\bibitem [{\citenamefont {Pertsova}\ \emph {et~al.}(2013)\citenamefont
  {Pertsova}, \citenamefont {Stamenova},\ and\ \citenamefont
  {Sanvito}}]{Pertsova13}%
  \BibitemOpen
  \bibfield  {author} {\bibinfo {author} {\bibfnamefont {A.}~\bibnamefont
  {Pertsova}}, \bibinfo {author} {\bibfnamefont {M.}~\bibnamefont
  {Stamenova}},\ and\ \bibinfo {author} {\bibfnamefont {S.}~\bibnamefont
  {Sanvito}},\ }\bibfield  {title} {\bibinfo {title} {Time-dependent electron
  transport through a strongly correlated quantum dot: multiple-probe
  open-boundary conditions approach},\ }\href
  {https://doi.org/10.1088/0953-8984/25/10/105501} {\bibfield  {journal}
  {\bibinfo  {journal} {J. Phys.: Condens. Matter}\ }\textbf {\bibinfo {volume}
  {25}},\ \bibinfo {pages} {105501} (\bibinfo {year} {2013})}\BibitemShut
  {NoStop}%
\bibitem [{\citenamefont {Vovchenko}\ \emph {et~al.}(2013)\citenamefont
  {Vovchenko}, \citenamefont {Anchishkin}, \citenamefont {Azema}, \citenamefont
  {Lombardo}, \citenamefont {Hayn},\ and\ \citenamefont
  {Daré}}]{Vovchenko_14}%
  \BibitemOpen
  \bibfield  {author} {\bibinfo {author} {\bibfnamefont {V.}~\bibnamefont
  {Vovchenko}}, \bibinfo {author} {\bibfnamefont {D.}~\bibnamefont
  {Anchishkin}}, \bibinfo {author} {\bibfnamefont {J.}~\bibnamefont {Azema}},
  \bibinfo {author} {\bibfnamefont {P.}~\bibnamefont {Lombardo}}, \bibinfo
  {author} {\bibfnamefont {R.}~\bibnamefont {Hayn}},\ and\ \bibinfo {author}
  {\bibfnamefont {A.-M.}\ \bibnamefont {Daré}},\ }\bibfield  {title} {\bibinfo
  {title} {{A new approach to time-dependent transport through an interacting
  quantum dot within the Keldysh formalism}},\ }\href
  {https://doi.org/10.1088/0953-8984/26/1/015306} {\bibfield  {journal}
  {\bibinfo  {journal} {J. Phys.: Condens. Matter}\ }\textbf {\bibinfo {volume}
  {26}},\ \bibinfo {pages} {015306} (\bibinfo {year} {2013})}\BibitemShut
  {NoStop}%
\bibitem [{\citenamefont {Altshuler}\ \emph {et~al.}(1982)\citenamefont
  {Altshuler}, \citenamefont {Aronov},\ and\ \citenamefont
  {Khmelnitsky}}]{Altshuler82}%
  \BibitemOpen
  \bibfield  {author} {\bibinfo {author} {\bibfnamefont {B.~L.}\ \bibnamefont
  {Altshuler}}, \bibinfo {author} {\bibfnamefont {A.~G.}\ \bibnamefont
  {Aronov}},\ and\ \bibinfo {author} {\bibfnamefont {D.~E.}\ \bibnamefont
  {Khmelnitsky}},\ }\bibfield  {title} {\bibinfo {title} {Effects of
  electron-electron collisions with small energy transfers on quantum
  localisation},\ }\href {https://doi.org/10.1088/0022-3719/15/36/018}
  {\bibfield  {journal} {\bibinfo  {journal} {J. Physics C}\ }\textbf {\bibinfo
  {volume} {15}},\ \bibinfo {pages} {7367} (\bibinfo {year}
  {1982})}\BibitemShut {NoStop}%
\bibitem [{\citenamefont {Marguerite}\ \emph {et~al.}(2016)\citenamefont
  {Marguerite}, \citenamefont {Cabart}, \citenamefont {Wahl}, \citenamefont
  {Roussel}, \citenamefont {Freulon}, \citenamefont {Ferraro}, \citenamefont
  {Grenier}, \citenamefont {Berroir}, \citenamefont
  {Pla\ifmmode~\mbox{\c{c}}\else \c{c}\fi{}ais}, \citenamefont {Jonckheere},
  \citenamefont {Rech}, \citenamefont {Martin}, \citenamefont {Degiovanni},
  \citenamefont {Cavanna}, \citenamefont {Jin},\ and\ \citenamefont
  {F\`eve}}]{Marguerite16}%
  \BibitemOpen
  \bibfield  {author} {\bibinfo {author} {\bibfnamefont {A.}~\bibnamefont
  {Marguerite}}, \bibinfo {author} {\bibfnamefont {C.}~\bibnamefont {Cabart}},
  \bibinfo {author} {\bibfnamefont {C.}~\bibnamefont {Wahl}}, \bibinfo {author}
  {\bibfnamefont {B.}~\bibnamefont {Roussel}}, \bibinfo {author} {\bibfnamefont
  {V.}~\bibnamefont {Freulon}}, \bibinfo {author} {\bibfnamefont
  {D.}~\bibnamefont {Ferraro}}, \bibinfo {author} {\bibfnamefont
  {C.}~\bibnamefont {Grenier}}, \bibinfo {author} {\bibfnamefont {J.-M.}\
  \bibnamefont {Berroir}}, \bibinfo {author} {\bibfnamefont {B.}~\bibnamefont
  {Pla\ifmmode~\mbox{\c{c}}\else \c{c}\fi{}ais}}, \bibinfo {author}
  {\bibfnamefont {T.}~\bibnamefont {Jonckheere}}, \bibinfo {author}
  {\bibfnamefont {J.}~\bibnamefont {Rech}}, \bibinfo {author} {\bibfnamefont
  {T.}~\bibnamefont {Martin}}, \bibinfo {author} {\bibfnamefont
  {P.}~\bibnamefont {Degiovanni}}, \bibinfo {author} {\bibfnamefont
  {A.}~\bibnamefont {Cavanna}}, \bibinfo {author} {\bibfnamefont
  {Y.}~\bibnamefont {Jin}},\ and\ \bibinfo {author} {\bibfnamefont
  {G.}~\bibnamefont {F\`eve}},\ }\bibfield  {title} {\bibinfo {title}
  {Decoherence and relaxation of a single electron in a one-dimensional
  conductor},\ }\href {https://doi.org/10.1103/PhysRevB.94.115311} {\bibfield
  {journal} {\bibinfo  {journal} {Phys. Rev. B}\ }\textbf {\bibinfo {volume}
  {94}},\ \bibinfo {pages} {115311} (\bibinfo {year} {2016})}\BibitemShut
  {NoStop}%
\bibitem [{\citenamefont {Jo}\ \emph {et~al.}(2022)\citenamefont {Jo},
  \citenamefont {Lee}, \citenamefont {Assouline}, \citenamefont {Brasseur},
  \citenamefont {Watanabe}, \citenamefont {Taniguchi}, \citenamefont {Roche},
  \citenamefont {Glattli}, \citenamefont {Kumada}, \citenamefont {Parmentier},
  \citenamefont {Sim},\ and\ \citenamefont {Roulleau}}]{jo22}%
  \BibitemOpen
  \bibfield  {author} {\bibinfo {author} {\bibfnamefont {M.}~\bibnamefont
  {Jo}}, \bibinfo {author} {\bibfnamefont {J.-Y.}\ \bibnamefont {Lee}},
  \bibinfo {author} {\bibfnamefont {A.}~\bibnamefont {Assouline}}, \bibinfo
  {author} {\bibfnamefont {P.}~\bibnamefont {Brasseur}}, \bibinfo {author}
  {\bibfnamefont {K.}~\bibnamefont {Watanabe}}, \bibinfo {author}
  {\bibfnamefont {T.}~\bibnamefont {Taniguchi}}, \bibinfo {author}
  {\bibfnamefont {P.}~\bibnamefont {Roche}}, \bibinfo {author} {\bibfnamefont
  {D.}~\bibnamefont {Glattli}}, \bibinfo {author} {\bibfnamefont
  {N.}~\bibnamefont {Kumada}}, \bibinfo {author} {\bibfnamefont
  {F.}~\bibnamefont {Parmentier}}, \bibinfo {author} {\bibfnamefont {H.~S.}\
  \bibnamefont {Sim}},\ and\ \bibinfo {author} {\bibfnamefont {P.}~\bibnamefont
  {Roulleau}},\ }\bibfield  {title} {\bibinfo {title} {{Scaling behavior of
  electron decoherence in a graphene Mach-Zehnder interferometer}},\ }\href
  {https://doi.org/10.1038/s41467-022-33078-2} {\bibfield  {journal} {\bibinfo
  {journal} {{Nat. Commun.}}\ }\textbf {\bibinfo {volume} {13}},\ \bibinfo
  {pages} {5473} (\bibinfo {year} {2022})}\BibitemShut {NoStop}%
\bibitem [{\citenamefont {N\'u\~nez Fern\'andez}\ \emph
  {et~al.}(2022)\citenamefont {N\'u\~nez Fern\'andez}, \citenamefont {Jeannin},
  \citenamefont {Dumitrescu}, \citenamefont {Kloss}, \citenamefont {Kaye},
  \citenamefont {Parcollet},\ and\ \citenamefont {Waintal}}]{Fernandez22}%
  \BibitemOpen
  \bibfield  {author} {\bibinfo {author} {\bibfnamefont {Y.}~\bibnamefont
  {N\'u\~nez Fern\'andez}}, \bibinfo {author} {\bibfnamefont {M.}~\bibnamefont
  {Jeannin}}, \bibinfo {author} {\bibfnamefont {P.~T.}\ \bibnamefont
  {Dumitrescu}}, \bibinfo {author} {\bibfnamefont {T.}~\bibnamefont {Kloss}},
  \bibinfo {author} {\bibfnamefont {J.}~\bibnamefont {Kaye}}, \bibinfo {author}
  {\bibfnamefont {O.}~\bibnamefont {Parcollet}},\ and\ \bibinfo {author}
  {\bibfnamefont {X.}~\bibnamefont {Waintal}},\ }\bibfield  {title} {\bibinfo
  {title} {{Learning Feynman Diagrams with Tensor Trains}},\ }\href
  {https://doi.org/10.1103/PhysRevX.12.041018} {\bibfield  {journal} {\bibinfo
  {journal} {Phys. Rev. X}\ }\textbf {\bibinfo {volume} {12}},\ \bibinfo
  {pages} {041018} (\bibinfo {year} {2022})}\BibitemShut {NoStop}%
\bibitem [{\citenamefont {Kloss}\ \emph {et~al.}(2018)\citenamefont {Kloss},
  \citenamefont {Weston},\ and\ \citenamefont {Waintal}}]{kloss18}%
  \BibitemOpen
  \bibfield  {author} {\bibinfo {author} {\bibfnamefont {T.}~\bibnamefont
  {Kloss}}, \bibinfo {author} {\bibfnamefont {J.}~\bibnamefont {Weston}},\ and\
  \bibinfo {author} {\bibfnamefont {X.}~\bibnamefont {Waintal}},\ }\bibfield
  {title} {\bibinfo {title} {{Transient and Sharvin resistances of Luttinger
  liquids}},\ }\href {https://doi.org/10.1103/PhysRevB.97.165134} {\bibfield
  {journal} {\bibinfo  {journal} {Phys. Rev. B}\ }\textbf {\bibinfo {volume}
  {97}},\ \bibinfo {pages} {165134} (\bibinfo {year} {2018})}\BibitemShut
  {NoStop}%
\bibitem [{sup(2025)}]{supplementary_data}%
  \BibitemOpen
  \bibfield  {title} {\bibinfo {title} {Supplementary data for manuscript},\
  }\href {https://doi.org/10.5281/zenodo.14864647} {10.5281/zenodo.14864647}
  (\bibinfo {year} {2025})\BibitemShut {NoStop}%
\bibitem [{\citenamefont {Waintal}\ \emph {et~al.}()\citenamefont {Waintal},
  \citenamefont {Wimmer}, \citenamefont {Akhmerov}, \citenamefont {Groth},
  \citenamefont {Nikolic}, \citenamefont {Istas}, \citenamefont {Örn
  Rosdahl},\ and\ \citenamefont {Varjas}}]{Waintal24}%
  \BibitemOpen
  \bibfield  {author} {\bibinfo {author} {\bibfnamefont {X.}~\bibnamefont
  {Waintal}}, \bibinfo {author} {\bibfnamefont {M.}~\bibnamefont {Wimmer}},
  \bibinfo {author} {\bibfnamefont {A.}~\bibnamefont {Akhmerov}}, \bibinfo
  {author} {\bibfnamefont {C.}~\bibnamefont {Groth}}, \bibinfo {author}
  {\bibfnamefont {B.~K.}\ \bibnamefont {Nikolic}}, \bibinfo {author}
  {\bibfnamefont {M.}~\bibnamefont {Istas}}, \bibinfo {author} {\bibfnamefont
  {T.}~\bibnamefont {Örn Rosdahl}},\ and\ \bibinfo {author} {\bibfnamefont
  {D.}~\bibnamefont {Varjas}},\ }\href {https://arxiv.org/abs/2407.16257}
  {\bibinfo {title} {Computational quantum transport}},\ \Eprint
  {https://arxiv.org/abs/2407.16257} {arXiv:2407.16257 [cond-mat.mes-hall]}
  \BibitemShut {NoStop}%
\bibitem [{\citenamefont {Groth}\ \emph {et~al.}(2014)\citenamefont {Groth},
  \citenamefont {Wimmer}, \citenamefont {Akhmerov},\ and\ \citenamefont
  {Waintal}}]{groth14}%
  \BibitemOpen
  \bibfield  {author} {\bibinfo {author} {\bibfnamefont {C.~W.}\ \bibnamefont
  {Groth}}, \bibinfo {author} {\bibfnamefont {M.}~\bibnamefont {Wimmer}},
  \bibinfo {author} {\bibfnamefont {A.~R.}\ \bibnamefont {Akhmerov}},\ and\
  \bibinfo {author} {\bibfnamefont {X.}~\bibnamefont {Waintal}},\ }\bibfield
  {title} {\bibinfo {title} {Kwant: a software package for quantum transport},\
  }\href {http://stacks.iop.org/1367-2630/16/i=6/a=063065} {\bibfield
  {journal} {\bibinfo  {journal} {New J. Phys.}\ }\textbf {\bibinfo {volume}
  {16}},\ \bibinfo {pages} {063065} (\bibinfo {year} {2014})}\BibitemShut
  {NoStop}%
\end{thebibliography}%

\end{document}